# STRONG DIRECT PRODUCT THEOREMS FOR QUANTUM COMMUNICATION AND QUERY COMPLEXITY

ALEXANDER A. SHERSTOV*

ABSTRACT. A *strong direct product theorem* (SDPT) states that solving $n$ instances of a problem requires $\Omega(n)$ times the resources for a single instance, even to achieve success probability $2^{-\Omega(n)}$. We prove that quantum communication complexity obeys an SDPT whenever the communication lower bound for a single instance is proved by the *generalized discrepancy method*, the strongest technique in that model. We prove that quantum query complexity obeys an SDPT whenever the query lower bound for a single instance is proved by the *polynomial method*, one of the two main techniques in that model. In both models, we prove the corresponding XOR lemmas and *threshold* direct product theorems.

## 1. INTRODUCTION

A natural question to ask of any computational model is how the resources needed to solve $n$ instances of a problem scale with $n$. More concretely, suppose that solving a single instance of a given decision problem, with probability of correctness $4/5$, requires $R$ units of a computational resource (such as time, memory, communication, or queries). How many units of the resource are needed to solve $n$ independent instances of the problem? Common sense suggests that the answer should be $\Omega(nR)$. After all, having less than $\epsilon nR$ units overall, for a small constant $\epsilon > 0$, leaves less than $\epsilon R$ units per instance, forcing the algorithm to guess random answers for many of the instances and resulting in overall success probability $2^{-\Omega(n)}$. Such a statement is called a *strong direct product theorem.* A related notion is an *XOR lemma*, which asserts that computing the XOR of the answers to the $n$ problem instances requires $\Omega(nR)$ resources, even if one is willing to settle for a success probability of $1/2 + 2^{-\Omega(n)}$. While highly plausible, XOR lemmas and strong direct product theorems are notoriously hard to prove and sometimes flat out wrong. To a considerable extent, the difficulty stems from the claimed exponential decay in the probability of successful computation. Dropping this part of the claim from strong direct product theorems results in *direct sum theorems*, which nevertheless are also elusive. The described nomenclature is fairly standard by now but does admit slight variations; for example, it may make more sense to work with success probability for an *average* instance rather than a *worst-case* instance, or vice versa.

Apart from their inherent importance in theoretical computer science, direct product-type results have various applications, including separations of circuit classes [28], improvement of soundness in proof systems [42], inapproximability results for optimization problems [11], [21], and time-space trade-offs [32], [5]. Perhaps the two most famous results in this line of research are Yao's XOR lemma [55] for circuits, which was in 1982 the first result of the kind, and Raz's parallel repetition theorem [42] for two-prover games. Considerable progress has been achieved in these and various other models, complemented by surprising counterexamples [20], [46], [8]. The models of interest to us in this paper are *quantum communication complexity* and *quantum query complexity*, where the direct product phenomenon is understood quite poorly. Furthermore, work here has advanced much more slowly than in the classical case, a point conveyed by the following overview of the classical and quantum literature.

* Microsoft Research, Cambridge, MA 02142. Email: sherstov@alumni.cs.utexas.edu .

*Classical communication and query complexity.* The direct sum problem in communication complexity was raised for the first time in the work of Karchmer, Raz, and Wigderson [28], who showed that its resolution for relations would yield an explicit function outside $\mathsf{NC}^1$. Feder, Kushilevitz, Naor, and Nisan [20] established a direct sum theorem for nondeterministic communication complexity and inferred a weaker result for deterministic communication. Information-theoretic methods have enabled substantial progress [16], [7], [24], [25], [22], [8] on the direct sum question in the randomized model and its restrictions, including one-way communication and simultaneous message passing. In what generality randomized communication complexity obeys a direct sum theorem remains unknown; some counterexamples have been discovered for a restricted choice of parameters [20].

It also remains unknown whether randomized communication complexity in general obeys a strong direct product theorem. A variety of results have been established, however, for concrete functions and some restrictions of the randomized model. Parnafes, Raz, and Wigderson [41] proved the first result of the kind, for "forests" of communication protocols. Shaltiel [46] proved an XOR lemma for uniform-distribution discrepancy, a well-studied communication complexity measure. Shaltiel's result has been generalized and strengthened in several ways [29], [10], [52], [37]. Jain, Klauck, and Nayak [23] obtained strong direct product theorems for an information-theoretic complexity measure called the *subdistribution bound*. Most recently, Klauck [31] proved the long-conjectured strong direct product theorem for the randomized communication complexity of the disjointness function.

In classical query complexity, the direct product phenomenon is well understood. Strong direct product theorems have been obtained for "decision forests" by Nisan, Rudich, and Saks [40], for "fair" decision trees by Shaltiel [46], and for the randomized query complexity of symmetric functions by Klauck, Špalek, and de Wolf [32]. Very recently, Drucker [18] obtained strong direct product theorems for the randomized query complexity of arbitrary functions.

*Quantum communication and query complexity.* Klauck, Špalek, and de Wolf [32] established a strong direct product theorem for the bounded-error quantum communication complexity of the disjointness function. Ben-Aroya, Regev, and de Wolf [12] proved that an analogous strong direct product theorem holds in the *one-way* quantum model. These are the only direct product theorems in quantum communication of which we are aware. The *direct sum* question has seen additional progress in the work of Jain, Radhakrishnan, and Sen [25] and Jain and Klauck [22], who examine one-way quantum communication and simultaneous message passing.

The results for quantum query complexity are just as few in number. The first direct product result is due to Aaronson [1], who proved it for the problem of $k$-fold search. Aaronson's result was improved to optimal with respect to all parameters by Klauck, Špalek, and de Wolf [32], who established a strong direct product theorem for the quantum query complexity of the OR function. In follow-up work, Ambainis, Špalek, and de Wolf [5] obtained a strong direct product theorem for all other symmetric functions. To our knowledge, these are the only direct product theorems in quantum query complexity. The only other result of which we are aware is due to Špalek [51], who developed a *multiplicative* version of the adversary method and proved that it obeys a strong direct product theorem.

**1.1. Our results.** In what follows, the symbol $f^{\otimes n}$ refers to the XOR of $n$ independent copies of a given decision problem $f$, which is a sign matrix in the case of communication complexity and a Boolean function $f\colon\{-1,+1\}^m \to \{-1,+1\}$ in the case of query complexity. The symbol $f^{(n)}$ refers to the task of simultaneously solving $n$ independent instances of $f$. In the latter context, we recall the notion of a *threshold* direct product theorem, which is a stronger statement than a *strong* direct product theorem. Specifically, a threshold direct product theorem defines successful computation of $f^{(n)}$ as correct computation of $(1-\beta)n$ instances for a small constant $\beta > 0$, as opposed to correct computation of all $n$ instances. A threshold direct product theorem states that computing $n$ instances requires $\Omega(n)$ times the resources for a single instance, even to achieve success probability $2^{-\Omega(n)}$ with this relaxed criterion of correct computation. All our direct product theorems are threshold direct product theorems.

*Quantum communication.* Let $\mathscr{R}$ denote the family of $0/1$ matrices in which the 1 entries form a submatrix. Such matrices are called *rectangles* and are the basic building blocks in communication complexity. In particular, the matrix $\Pi$ of acceptance probabilities of any communication protocol with cost $c$ obeys

$$\Pi \in 2^{O(c)} \operatorname{conv}\{\pm R : R \in \mathscr{R}\}. \tag{1.1}$$

This fact has an elementary and well-known demonstration [35] for classical protocols. The validity of (1.1) for quantum protocols, on the other hand, was open for several years and settled relatively recently in an elegant paper of Linial and Shraibman [39]. This fact immediately gives a criterion for high communication complexity, known as the *generalized discrepancy method*. Specifically, define a norm $\mu$ on matrices by letting $\mu(\Pi)$ be the least $K \geqslant 0$ for which $\Pi \in K \operatorname{conv}\{\pm R : R \in \mathscr{R}\}$. Then a sign matrix has high bounded-error communication complexity whenever every real matrix in its neighborhood has high $\mu$ norm. The method has an equivalent *dual* formulation that is widely used and has a rich history, e.g., [30], [43], [39], [48], [49], [36].

Linial and Shraibman [39] showed that the generalized discrepancy method subsumes all earlier criteria for high quantum communication complexity. In particular, all known lower bounds for two-way quantum communication can be derived using the generalized discrepancy method and no additional facts about quantum communication. Furthermore, the full power of the method is rarely necessary, and the main lower bounds have all been obtained using a simpler criterion known as the trace norm method, e.g., [56], [34], [30], [43], [48], [50].

Our main result is that the generalized discrepancy method obeys an XOR lemma and a threshold direct product theorem. This solves an open problem posed in [37, Sec. 6]. In particular, whenever the generalized discrepancy method yields a tight lower bound on the quantum communication complexity of a sign matrix $F$ (as it does for all known $F$), one immediately obtains an XOR lemma and threshold direct product theorem for $F$. In what follows, we let the real number $\mathrm{GDM}_\epsilon(F)$ denote the lower bound that the generalized discrepancy method gives on the $\epsilon$-error quantum communication complexity of $F$.

THEOREM 1.1. *Fix a sign matrix $F$. Then the following tasks require $\Omega(n\,\mathrm{GDM}_{1/5}(F))$ qubits of communication each:*

- *solving $F^{\otimes n}$ with worst-case probability $1/2 + 2^{-\Omega(n)}$;*
- *solving $F^{(n)}$ with worst-case probability $2^{-\Omega(n)}$.*

*The same holds for solving with probability $2^{-\Omega(n)}$ at least $(1-\beta)n$ among $n$ instances of $F$, for small $\beta > 0$.*

It is natural to consider the direct product question in the broader context of distinct communication problems $F_1, F_2, \dots, F_n$, rather than $n$ instances of the same communication problem. This paper gives a detailed solution in the generalized setting as well. As before, we consider the task of computing the XOR of the answers, denoted $F_1 \otimes \cdots \otimes F_n$, and the task of solving each of the $n$ problems, denoted $(F_1, \dots, F_n)$. Here, one clearly cannot hope to prove that $\Omega(\sum_{i=1}^n \mathrm{GDM}_{1/5}(F_i))$ is a communication lower bound for solving the above two tasks with advantage $2^{-\Omega(n)}$ over random guessing. Indeed, if $F_1$ has communication cost larger than the other problems combined, then for all intents and purposes we are working with a *single* problem, and no exponential decay in success probability is possible by definition. However, it is reasonable to expect a direct sum theorem here—and we prove that it indeed holds:

THEOREM 1.2. *For all sign matrices $F_1, F_2, \dots, F_n$ of rank $>1$, computing $F_1 \otimes \cdots \otimes F_n$ with probability $4/5$ requires a communication protocol with cost $\Omega\left(\sum_{i=1}^n \mathrm{GDM}_{1/5}(F_i)\right)$.*

We complement Theorem 1.2 by proving that a quantum protocol's success probability does indeed become exponentially close to that of random guessing when the protocol's communication is bounded by the sum of the smallest $\lceil 0.99n \rceil$ of the numbers $\mathrm{GDM}_{1/5}(F_1), \dots, \mathrm{GDM}_{1/5}(F_n)$:

THEOREM 1.3. *Fix sign matrices $F_1, F_2, \dots, F_n$ of rank greater than 1. Then the following tasks have quantum communication cost $\Omega\left(\min_{|S|=\lceil 0.99n \rceil} \left\{\sum_{i\in S} \mathrm{GDM}_{1/5}(F_i)\right\}\right)$:*

- *solving $F_1 \otimes \cdots \otimes F_n$ with worst-case probability $1/2 + 2^{-\Omega(n)}$;*
- *solving $(F_1, \dots, F_n)$ with worst-case probability $2^{-\Omega(n)}$.*

*The same holds for solving with probability $2^{-\Omega(n)}$ at least $(1-\beta)n$ among the $n$ instances, for small $\beta > 0$.*

All the theorems above are valid for quantum protocols with arbitrary prior entanglement. While stated above for *worst-case* complexity, Theorems 1.1–1.3 hold for average-case complexity under a certain joint probability distribution defined explicitly in our proof. Finally, we prove results identical to Theorems 1.1–1.3 in the setting of *partial* communication problems, whose domain of definition is a proper subset of all possible inputs. In such cases $\mathrm{GDM}_\epsilon(F)$ is computed by considering the smallest $\mu$ norm over real matrices whose entries are within $\epsilon$ of the values of $F$ on the domain of $F$ and range unrestricted in $[-1-\epsilon, 1+\epsilon]$ outside the domain of $F$.

*Quantum query complexity.* The polynomial method, discovered by Beals, Buhrman, Cleve, Mosca, and de Wolf [9], is a technique for proving lower bounds on quantum query complexity. It is easy to state: The acceptance probability of a quantum query algorithm on input $x \in \{-1,+1\}^m$ is a real polynomial in $x_1, x_2, \dots, x_m$ of degree at most $2T$, where $T$ is the number of queries. Conversely, if there is no degree-$d$ real polynomial that approximates a given Boolean function $f$ within $1/5$ on all inputs, then the $1/5$-error query complexity of $f$ is at least $d/2$. Beals et al. [9] used this method to obtain tight lower bounds on the query complexity of all symmetric functions. The polynomial method has since yielded many other tight lower bounds, e.g., [13], [2], [1], [32]. The main alternative to the polynomial method is the *adversary method*, introduced by Ambainis [4] and augmented in subsequent works, including a multiplicative version of the method due to Špalek [51]. The polynomial method and adversary method are incomparable: functions are known on which either technique outperforms the other.

Our second main result is that the polynomial method obeys an XOR lemma and a threshold direct product theorem. In particular, whenever the polynomial method yields a tight lower bound on the query complexity of a given Boolean function $f$, one automatically obtains an XOR lemma and a threshold direct product theorem for $f$. This subsumes the functions $f$ in all previous direct product theorems for quantum query complexity [1], [32], [5]. As for communication complexity, we prove our results in the general setting of distinct functions $f_1, f_2, \dots, f_n$ rather than $n$ instances of the same function $f$. In the statement to follow, the symbol $\deg_\epsilon(f)$ stands for the least degree of a real polynomial that approximates $f$ within $\epsilon$ pointwise.

THEOREM 1.4. *For all functions $f_1, f_2, \dots, f_n\colon \{-1,+1\}^m \to \{-1,+1\}$, computing $f_1 \otimes \cdots \otimes f_n$ with probability $4/5$ requires a quantum query algorithm with cost $\Omega\big(\sum_{i=1}^n \deg_{1/5}(f_i)\big)$.*

We complement this with a direct product result analogous the one for communication:

THEOREM 1.5. *Fix functions $f_1, f_2, \dots, f_n\colon \{-1,+1\}^m \to \{-1,+1\}$. Then the following tasks require $\Omega\big(\min_{|S|=\lceil 0.99n\rceil} \big\{\sum_{i\in S} \deg_{1/5}(f_i)\big\}\big)$ quantum queries each:*

- *solving $f_1 \otimes \cdots \otimes f_n$ with worst-case probability $1/2 + 2^{-\Omega(n)}$;*
- *solving $(f_1, \dots, f_n)$ with worst-case probability $2^{-\Omega(n)}$.*

*The same holds for solving with probability $2^{-\Omega(n)}$ at least $(1-\beta)n$ among the $n$ instances, for small $\beta > 0$.*

In particular, Theorem 1.5 shows that for every Boolean function $f$, the tasks of computing $f^{\otimes n}$ and $f^{(n)}$ each have quantum query complexity $\Omega(n \deg_{1/5}(f))$, even to achieve advantage $2^{-\Omega(n)}$ over random guessing. The additional remarks made earlier in the context of communication carry over in full. Specifically, Theorems 1.4 and 1.5 remain valid for *partial* Boolean functions, in which case the approximate degree $\deg_\epsilon(f)$ is defined as the least degree of a polynomial that approximates $f$ within $\epsilon$ on the domain of $f$ and ranges freely in $[-1-\epsilon, 1+\epsilon]$ everywhere else on the hypercube. Lastly, Theorems 1.4 and 1.5 are stated *worst-case* complexity but are also valid for average-case complexity under a certain joint probability distribution given explicitly in our proof.

*Consequences for polynomial approximation.* In proving Theorem 1.5, we show in particular that

$$\deg_{1-2^{-\Omega(n)}}(f^{\otimes n}) \geqslant \Omega(n \deg_{1/5}(f)).$$

To our knowledge, this is the first direct product theorem for polynomial approximation. It is optimal by an upper bound due to Buhrman, Newman, Röhrig, and de Wolf [14], who proved that $\deg_{1/5}(f^{\otimes n}) = O(n \deg_{1/5}(f))$. In Section 6, we also obtain the first *direct sum* results for polynomial approximation: given any function $F\colon \{-1,+1\}^n \to \{-1,+1\}$ with $\deg_{1/5}(F) = \Omega(n)$ (which includes the familiar majority and parity functions and the random functions), we prove that

$$\deg_{1/5}(F(f, f, \ldots, f)) = \Omega(\deg_{1/5}(F) \deg_{1/5}(f)),$$

for all Boolean functions $f$. This lower bound matches the upper bound in [14].

**1.2. Our techniques.** The proof technique of this paper is quite general and applies to any bounded-error model of computation that admits a representation as a *convex subset* of a real linear space. Examples of convex subsets that naturally arise from a computational model include the unit ball of a norm and the linear span of a given set of functions. Both of these cases are treated in this paper: the former corresponds to communication complexity and the latter, to query complexity. For simplicity, we will focus on the former setting in this overview.

Here, one fixes a finite set $X$ and lets the space of real functions on $X$ and $X^n$ be normed by $|||\cdot|||$. The norm captures the complexity of exact computation, as measured in the relevant resource. In other words, functions that represent low-cost communication protocols and low-cost query algorithms will have small norm. The complexity of $\epsilon$-error computation for a given function $f\colon X \to \{-1,+1\}$ is then given by the minimum norm in the $\epsilon$-neighborhood of $f$. This norm-based formalism is particularly natural in quantum computing and has been in use for many years, e.g., [56], [34], [43].

XOR lemmas for *correlation* represent a particularly well-studied form of hardness amplification in this setting: given a function $f\colon X \to \{-1,+1\}$ that has small correlation $\gamma$ with all low-cost communication protocols or low-cost query algorithms, one argues that for $f^{\otimes n}$ the correlation further drops to $\gamma^{\Omega(n)}$. In the language of norms, a function $f$ has small correlation with the simple functions if and only if the dual norm $|||f|||^*$ is small. Thus, an XOR lemma for correlation is an assertion about the multiplicativity of the dual norm: $|||f^{\otimes n}|||^* \leqslant (|||f|||^*)^{\Omega(n)}$. Much of the research surveyed above fits in this framework, including [46], [17], [52], [37].

This paper addresses a rather different problem. While we also seek to establish XOR lemmas, we start with a much more general object: a function $f\colon X \to \{-1,+1\}$ with high bounded-error computational complexity. The key point is that $f$ need no longer have small dual norm $|||f|||^*$, or equivalently small correction with the low-cost functions. Indeed, many common functions with near-maximum bounded-error complexity, such as the OR function in query complexity and the disjointness function in communication complexity, have high correlation with the low-cost protocols and query algorithms under *every* distribution on the domain. In particular, the above research on XOR lemmas for correlation no longer applies.

This described difficulty crystallizes best in the language of norms. By duality, a given function $f\colon X \to \{-1,+1\}$ of interest has high bounded-error complexity if and only if there exists a real-valued function $\psi\colon X \to \{-1,+1\}$ of unit $\ell_1$ norm that has reasonably large inner product $\langle f, \psi\rangle$ but low dual norm $|||\psi|||^*$. This function $\psi$ is a *witness* to the fact that $f$ has high bounded-error complexity. The challenge is to construct a corresponding witness for $f^{\otimes n}$. The natural candidate, $\psi^{\otimes n}$, is perfectly useless for this purpose: while $\|\psi^{\otimes n}\|_1 = 1$ and moreover we can certainly hope for an exponential decay in the dual norm $|||\psi^{\otimes n}|||^* \leqslant (|||\psi|||^*)^{\Omega(n)}$, the correlation with $f^{\otimes n}$ will also decay exponentially: $\langle f^{\otimes}, \psi^{\otimes n}\rangle = \langle f, \psi\rangle^n$. This translates to an uninteresting statement like "computing $f^{\otimes n}$ with error probability $2^{-\Omega(n)}$ incurs $\Omega(n)$ times the cost of computing $f$." What we want is just the opposite: the error probability allowed in computing $f^{\otimes n}$ should be exponentially close to the trivial rate $1/2$ rather than to 0.

The crux of our solution is the construction of the sought witness for $f^{\otimes n}$, using ideas from approximation theory to design a joint, nonproduct distribution under which the correlation of $\psi^{\otimes}$ and $f^{\otimes}$ becomes extremely high but the dual norm $|||\psi^{\otimes}|||$ remains extremely low. This construction works for any norm whose dual $|||\cdot|||^*$ possesses a multiplicative property, as our norm of interest for which multiplicativity was established in previous work by Cleve, Slofstra, Unger, and Upadhyay [17].

This sketches some ideas in the proofs of the XOR lemmas. The direct product theorems are then derived by combining the XOR lemmas with known results on the low-error approximation of symmetric Boolean functions. In particular, we appeal to a result of de Wolf [54] that OR and other symmetric Boolean functions admit uniform approximation to within $2^{-\Omega(n)}$ by a polynomial of reasonably small degree, $O(n)$.

**1.3. Organization.** The remainder of this manuscript is organized as follows. Section 2 opens with a review of technical preliminaries. Our generic method for proving XOR lemmas and direct product theorems for bounded-error computation is developed in Section 3. Applications to communication complexity and query complexity are presented in Sections 4 and 5. Further generalizations in the context of polynomial approximation are provided in Section 6.

## 2. Preliminaries

We view Boolean functions as mappings $f\colon X \to \{-1,+1\}$ for some finite set $X$, where $-1$ and $+1$ correspond to "true" and "false," respectively. A *partial* Boolean function $g$ on a finite set $X$ is a mapping $g\colon D \to \{-1,+1\}$ for some nonempty proper subset $D \subset X$.

We denote the domain of $g$ by $\operatorname{dom} g = D$. For emphasis, we will occasionally refer to Boolean functions with $\operatorname{dom} g = X$ as *total.* For a string $x \in \{-1,+1\}^n$, we use the shorthand $|x| = |\{i : x_i = -1\}| = \sum(1-x_i)/2$. For $\epsilon_1, \epsilon_2, \dots, \epsilon_n \in [0,1]$, the symbol $\Pi(\epsilon_1, \epsilon_2, \dots, \epsilon_n)$ stands for the product distribution on $\{-1,+1\}^n$ whereby the $i$th bit of the string takes on $-1$ with probability $\epsilon_i$, independently for each $i$. For an event $E$, the corresponding indicator function is

$$\mathbf{I}[E] = \begin{cases} 1 & \text{if } E \text{ holds,} \\ 0 & \text{otherwise.} \end{cases}$$

We adopt the standard definition of the sign function:

$$\operatorname{sgn} t = \begin{cases} -1, & t < 0, \\ 0, & t = 0, \\ 1, & t > 0. \end{cases}$$

It will also be convenient to define a modified sign function,

$$\widetilde{\operatorname{sgn}}\, t = \begin{cases} -1, & t < 0, \\ 1, & t \geqslant 0. \end{cases}$$

We will specify an $n$-bit string by its $i$th bit, for example, $(\dots, (\epsilon - \epsilon_i)/(1-\epsilon_i), \dots)$ or $(\dots, z_i, \dots)$. Given a function $\phi\colon \{-1,+1\}^n \to \mathbb{R}$, there exists a unique *multilinear* polynomial $\tilde{\phi}\colon \mathbb{R}^n \to \mathbb{R}$ such that $\phi \equiv \tilde{\phi}$ on $\{-1,+1\}^n$. We will always identify $\phi$ with its multilinear extension $\tilde{\phi}$ to $\mathbb{R}^n$. In particular, we will write $\phi(z)$ for arbitrary $z \in [-1,1]^n$. This convention requires some care. To illustrate, consider the polynomial $p_k\colon \{-1,+1\}^n \to \mathbb{R}$ from Lemma 3.1, defined by

$$p_k(x) = (-1)^k \prod_{i=1}^{k} (|x| - i).$$

One may be tempted to evaluate the multilinear extension of $p_k$ to $\mathbb{R}^n$ by direct substitution. This would of course be incorrect because the defining equation for $p_k$ was only meant to be valid on the hypercube and is not multilinear.

For integers $n \geqslant k \geqslant 0$, we adopt the shorthand

$$\binom{n}{\leqslant k} = \binom{n}{0} + \binom{n}{1} + \cdots + \binom{n}{k}.$$

Throughout this manuscript, $\log x$ stands for the logarithm of $x$ to the base 2. The binary entropy function $H\colon [0,1] \to [0,1]$ is given by $H(p) = -p \log p - (1-p)\log(1-p)$ and is strictly increasing on $[0,1/2]$. The following bound is well known [26, p. 283]:

$$\binom{n}{\leqslant k} \leqslant 2^{H(k/n)n}, \qquad\qquad k = 0,1,2,\dots,\lfloor n/2 \rfloor.$$

The Cartesian product of sets $X_1, X_2, \dots, X_n$ is denoted $\prod X_i$, or for greater explicitness $X_1 \times X_2 \times \cdots \times X_n$. The degree of a real polynomial $p$ is denoted $\deg p$.

**2.1. Norms and duality.** For a finite set $X$, the linear space of real functions on $X$ is denoted $\mathbb{R}^X$. This space is equipped with the usual norms and inner product:

$$\left.\begin{aligned} \|\phi\|_\infty &= \max_{x\in X} |\phi(x)| && (\phi \in \mathbb{R}^X), \\ \|\phi\|_1 &= \sum_{x\in X} |\phi(x)| && (\phi \in \mathbb{R}^X), \\ \langle \phi, \psi\rangle &= \sum_{x\in X} \phi(x)\psi(x) && (\phi, \psi \in \mathbb{R}^X). \end{aligned}\right\} \tag{2.1}$$

The tensor product of $\phi \in \mathbb{R}^X$ and $\psi \in \mathbb{R}^Y$ is the function $\phi \otimes \psi \in \mathbb{R}^{X\times Y}$ given by $(\phi \otimes \psi)(x, y) = \phi(x)\psi(y)$. The tensor product $\phi \otimes \phi \otimes \cdots \otimes \phi$ ($n$ times) is denoted $\phi^{\otimes n} \in \mathbb{R}^{X^n}$. When specialized to real matrices, the tensor product is the usual Kronecker product. The pointwise (Hadamard) product of $\phi, \psi \in \mathbb{R}^X$ is denoted $\phi \circ \psi \in \mathbb{R}^X$ and given by $(\phi \circ \psi)(x) = \phi(x)\psi(x)$. Note the difference between $\phi \otimes \psi$ and $\phi \circ \psi$.

For an arbitrary norm $|||\cdot|||$ on $\mathbb{R}^X$, recall that $|||\cdot|||^*$ refers to the dual norm given by $|||\phi|||^* = \max_{\psi\neq 0}\langle \phi, \psi\rangle / |||\psi|||$. A corollary to the duality $|||\cdot|||^{**} = |||\cdot|||$ is the following classical fact pertaining to approximation; cf. [36, Thm. 6.3].

FACT 2.1. *Let $X$ be a finite set, let $N_1, N_2$ be norms on $\mathbb{R}^X$. Then for any $\epsilon \geqslant 0$ and any $f \in \mathbb{R}^X$ with $N_2(f) > \epsilon$,*

$$\min\{N_1(f - \xi) : N_2(\xi) \leqslant \epsilon\} = \max_{\psi\neq 0} \frac{\langle f, \psi\rangle - \epsilon N_2^*(\psi)}{N_1^*(\psi)}. \tag{2.2}$$

*Proof.* One can restate (2.2) as

$$\sup\{c \geqslant 0 : f + \epsilon B_{N_2} \cap c B_{N_1} = \varnothing\} = \max_{\psi\neq 0} \frac{\langle f, \psi\rangle - \epsilon N_2^*(\psi)}{N_1^*(\psi)}, \tag{2.3}$$

where $B_{N_1}$ and $B_{N_2}$ stand for the unit balls of $N_1$ and $N_2$, respectively. By the separating hyperplane theorem, the compact convex sets $f + \epsilon B_{N_2}$ and $c B_{N_1}$ are disjoint if and only if $\langle \psi, f + \zeta_2\rangle > \langle \psi, \zeta_1\rangle$ for some $\psi \in \mathbb{R}^X$ and all $\zeta_1 \in c B_{N_1}, \zeta_2 \in \epsilon B_{N_2}$, which by duality is equivalent to $\langle f, \psi\rangle - \epsilon N_2^*(\psi) > c N_1^*(\psi)$. This forces equality in (2.3). ∎

We will be mainly concerned with approximation in the infinity norm. This case is served by the notation $|||f|||_\epsilon = \min\{|||f - \xi||| : \|\xi\|_\infty \leqslant \epsilon\}$, where $f : X \to \mathbb{R}$ is a given function and $|||\cdot|||$ is an arbitrary norm on $\mathbb{R}^X$. When $f$ is a partial Boolean function on $X$, we define $|||f|||_\epsilon$ to be the least norm $|||\phi|||$ over all elements $\phi \in \mathbb{R}^X$ such

$$\begin{aligned} |f(x) - \phi(x)| &\leqslant \epsilon, && x \in \operatorname{dom} f, \\ |\phi(x)| &\leqslant 1 + \epsilon, && x \notin \operatorname{dom} f. \end{aligned}$$

When $\operatorname{dom} f = X$, this agrees with the earlier definition of the symbol $|||f|||_\epsilon$. A straightforward consequence of Fact 2.1 is:

COROLLARY 2.2. *Let $X$ be a finite set, $|||\cdot|||$ a norm on $\mathbb{R}^X$. Then for every $\epsilon \in (0,1)$ and every* (*possibly partial*) *Boolean function $f$ on $X$,*

$$|||f|||_\epsilon = \max_{\psi \in \mathbb{R}^X \setminus \{0\}} \frac{1}{|||\psi|||^*} \left\{ \sum_{x \in \operatorname{dom} f} f(x)\psi(x) - \sum_{x \notin \operatorname{dom} f} |\psi(x)| - \epsilon \|\psi\|_1 \right\} \tag{2.4}$$

$$\geqslant \max_{\psi \in \mathbb{R}^X \setminus \{0\}} \frac{1}{|||\psi|||^*} \left\{ 2 \sum_{x \in \operatorname{dom} f} f(x)\psi(x) - (1+\epsilon)\|\psi\|_1 \right\}. \tag{2.5}$$

*In particular, when* $\operatorname{dom} f = X$,

$$|||f|||_\epsilon = \max_{\psi \in \mathbb{R}^X \setminus \{0\}} \frac{\langle f, \psi \rangle - \epsilon \|\psi\|_1}{|||\psi|||^*}.$$

*Proof.* Let $\mathbb{R}^X$ be normed by $N_1 = |||\cdot|||$ and

$$N_2(\psi) = \max \left\{ \max_{x \in \operatorname{dom} f} |\psi(x)|, \frac{\epsilon}{1+\epsilon} \max_{x \notin \operatorname{dom} f} |\psi(x)| \right\}.$$

Then $|||f|||_\epsilon = \min\{N_1(F - \xi) : N_2(\xi) \leqslant \epsilon\}$, where $F$ is the extension of $f$ to $X$ given by $F = 0$ outside $\operatorname{dom} f$. Since $N_2^*(\psi) = \|\psi\|_1 + \sum_{x \notin \operatorname{dom} f} |\psi(x)|/\epsilon$, Fact 2.1 implies (2.4). Finally, (2.5) is immediate from (2.4). ∎

**2.2. Matrix analysis.** A special case covered by the notation (2.1) is the family $\mathbb{R}^{n \times m}$ of all real matrices of dimension $n \times m$. More explicitly, one has $\|A\|_\infty = \max |A_{ij}|$, $\|A\|_1 = \sum |A_{ij}|$, and $\langle A, B \rangle = \sum A_{ij} B_{ij}$ for all $A, B \in \mathbb{R}^{n \times m}$. For finite sets $X$ and $Y$, we let $\mathbb{R}^{X \times Y}$ and $\{-1,+1\}^{X \times Y}$ stand for the families of real and $\pm 1$ matrices, respectively, with rows indexed by elements of $X$ and columns indexed by elements of $Y$. The rank of a matrix $A$ over the reals is denoted $\operatorname{rk} A$. The symbol $\operatorname{diag}(a_1, a_2, \ldots, a_n)$ refers to the diagonal matrix of order $n$ with entries $a_1, a_2, \ldots, a_n$ on the diagonal. A *signature scaling* of a matrix $M \in \mathbb{R}^{n \times m}$ is any matrix of the form $\operatorname{diag}(a_1, \ldots, a_n) M \operatorname{diag}(b_1, \ldots, b_m)$, where $a_1, \ldots, a_n, b_1, \ldots, b_m \in \{-1, +1\}$. The symbols $I_n$ and $J_{n,m}$ refer to the identity matrix of order $n$ and the all-ones matrix of dimension $n \times m$, respectively; we will drop the subscripts and write simply $I, J$ whenever the dimension is clear from the context. A *sign matrix* is any matrix with entries $\pm 1$. A *Hadamard matrix* is any sign matrix $A$ of order $n$ that obeys $AA^{\mathsf{T}} = nI$. The property of being a Hadamard matrix is preserved under signature scaling. In particular, the Hadamard matrices of order 2 are precisely the signature scalings of the matrix

$$H = \begin{bmatrix} 1 & 1 \\ 1 & -1 \end{bmatrix}. \tag{2.6}$$

Recall that the tensor product of Hadamard matrices is a Hadamard matrix. In particular, $H^{\otimes n}$ is a Hadamard matrix for every integer $n \geqslant 1$, a classical construction due to J. J. Sylvester.

The Frobenius norm of $M \in \mathbb{R}^{n \times m}$ is given by $\|M\|_{\mathrm{F}} = \sqrt{\sum M_{ij}^2}$. The singular values of $M$ are denoted $\sigma_1(M) \geqslant \sigma_2(M) \geqslant \sigma_3(M) \geqslant \cdots$, with the spectral norm and trace norm of $M$ given by $\|M\| = \sigma_1(M)$ and $\|M\|_\Sigma = \sum \sigma_i(M)$, respectively. The spectral norm and trace norm are duals. An equivalent definition of the trace norm is

$$\|M\|_\Sigma = \min\{\|A\|_{\mathrm{F}}\|B\|_{\mathrm{F}} : AB = M\}. \tag{2.7}$$

A close relative of the trace norm is the $\gamma_2$ norm, defined by

$$\gamma_2(M) = \min\{\|A\|_{\mathrm{row}}\|B\|_{\mathrm{col}} : AB = M\}, \tag{2.8}$$

where $\|A\|_{\mathrm{row}}$ and $\|B\|_{\mathrm{col}}$ stand for the largest Euclidean norm of a row of $A$ and the largest Euclidean norm of a column of $B$, respectively. The subscript in $\gamma_2$ is a reference to the norm used to measure the rows of $A$ and columns of $B$, namely, the Euclidean norm $\ell_2$. Put

$$\gamma_{2,\epsilon}(M) = \min\{\gamma_2(M - E) : \|E\|_\infty \leqslant \epsilon\},$$

the least $\gamma_2$ norm of a matrix in the $\epsilon$-neighborhood of $M$. While $\gamma_2$, the trace norm, spectral norm, and Frobenius norm are all norms, only the latter three are *matrix* norms. Note further that $\gamma_{2,\epsilon}$ is not a norm (homogeneity fails). We collect some well-known properties of $\gamma_2$ in the following statement.

FACT 2.3. *Let $A$ and $B$ denote arbitrary real matrices. Let $H_N$ be a Hadamard matrix of order $N$. Then:*

(i) $\gamma_2(A) = \gamma_2(B)$ *whenever $A$ is a signature scaling of $B$,*
(ii) $\gamma_2(A) \geqslant \gamma_2(B)$ *whenever $B$ is a submatrix of $A$,*
(iii) $\gamma_2$ *is invariant under duplication of rows or columns,*
(iv) $\gamma_2(A) \geqslant \|A\|_\infty$,
(v) $\gamma_2(A) \geqslant \|A\|_\Sigma/\sqrt{nm}$ *for all* $A \in \mathbb{R}^{n\times m}$,
(vi) $\gamma_2^*(A) \leqslant \|A\|\sqrt{nm}$ *for all* $A \in \mathbb{R}^{n\times m}$,
(vii) $\gamma_2(J) = 1$,
(viii) $\gamma_2(H_N) = \sqrt{N}$,
(ix) $\gamma_{2,\epsilon}(A) \geqslant (1-\epsilon)\sqrt{nm}/\|A\|$ *for every* $A \in \{-1,+1\}^{n\times m}$,
(x) $\gamma_{2,\epsilon}(J) = 1 - \epsilon$ *for* $0 \leqslant \epsilon \leqslant 1$,
(xi) $\gamma_{2,\epsilon}(H_N) = (1-\epsilon)\sqrt{N}$ *for* $0 \leqslant \epsilon \leqslant 1$,
(xii) $\gamma_2(A \otimes B) \leqslant \gamma_2(A)\gamma_2(B)$,
(xiii) $\gamma_2(A \circ B) \leqslant \gamma_2(A)\gamma_2(B)$.

Tracing the authorship of the items in Fact 2.3 is somewhat challenging. Items (v) and (vi) appear in [38], and the others are likely classical. For the reader's convenience, we include the short proofs for all.

*Proof.* (i)–(iii) Immediate from the definition (2.8).

(iv) By (2.8), one has $A_{ij} = \langle u_i, v_j\rangle$ for some vectors $u_i, v_j$ with $\|u_i\|_2\|v_j\|_2 \leqslant \gamma_2(A)$, whence $|A_{ij}| \leqslant \gamma_2(A)$ by the Cauchy-Schwarz inequality.
(v) Immediate from the definitions (2.7), (2.8).
(vi) Equivalent to (v) by duality.
(vii) The upper bound follows from (2.8) and $J_{n,m} = J_{n,1}J_{1,m}$. The lower bound follows from (iv).

(viii) The upper bound follows from (2.8) and $H_N = H_N I$. The lower bound follows from (v) and $\|H_N\|_\Sigma = N\sqrt{N}$.

(ix) Take $|||\cdot||| = \gamma_2$, $f = A$, and $\psi = \frac{1}{nm}A$ in Corollary 2.2 and apply (vi).

(x) $M = (1-\epsilon)J$ obeys $\gamma_2(M) = 1-\epsilon$ by (vii) and $\|J - M\|_\infty = \epsilon$, which gives the upper bound. The lower bound follows from (ix).

(xi) $M = (1-\epsilon)H_N$ obeys $\gamma_2(M) = (1-\epsilon)\sqrt{N}$ by (viii) and $\|H_N - M\|_\infty = \epsilon$, which gives the upper bound. The lower bound follows from (ix) and $\|H_N\| = \sqrt{N}$. ∎

(xii) Immediate from the definition (2.8) and the so-called *mixed-product property*: if $A = XY$ and $B = X'Y'$, then $A \otimes B = (X \otimes X')(Y \otimes Y')$.

(xiii) Immediate from (xii) in view of (ii) and the fact that $A \circ B$ is a submatrix of $A \otimes B$.

In the context of lower bounds on communication complexity, we will encounter *partial* sign matrices, which are matrices with entries in $\{-1, +1, *\}$. For a partial sign matrix $F$ and a norm $|||\cdot|||$, we let $|||F|||_\epsilon$ stand for the least norm $|||M|||$ of a real matrix $M$ with

$$\begin{aligned} |F_{ij} - M_{ij}| &\leqslant \epsilon && (F_{ij} = \pm 1),\\ |M_{ij}| &\leqslant 1+\epsilon && (F_{ij} = *). \end{aligned}$$

This is an instantiation for matrices of an earlier definition. The primary case of interest to us will be $\gamma_{2,\epsilon}(F)$.

**2.3. Approximation by polynomials.** Let $f\colon X \to \{-1,+1\}$ be given, for a finite subset $X \subset \mathbb{R}^n$. The *$\epsilon$-approximate degree* of $f$, denoted $\deg_\epsilon(f)$, is the least degree of a real polynomial $p$ such that $\|f - p\|_\infty \leqslant \epsilon$. We generalize this definition to a partial Boolean function $f$ on $X$ by letting $\deg_\epsilon(f)$ be the least degree of a real polynomial $p$ with

$$\begin{aligned} |f(x) - p(x)| &\leqslant \epsilon, && x \in \operatorname{dom} f,\\ |p(x)| &\leqslant 1+\epsilon, && x \in X \setminus \operatorname{dom} f. \end{aligned}$$

By basic approximation theory [19], there is a univariate polynomial of degree $O(\log\frac{1}{\epsilon})$ that sends $[-\frac{4}{3}, \frac{4}{3}] \to [-1-\epsilon, 1+\epsilon]$, $[-\frac{4}{3}, -\frac{2}{3}] \to [-1-\epsilon, -1+\epsilon]$ and $[\frac{2}{3}, \frac{4}{3}] \to [1-\epsilon, 1+\epsilon]$. Thus, every (possibly partial) Boolean function $f$ obeys

$$\deg_\epsilon(f) \leqslant O\left(\deg_{1/3}(f)\log\frac{1}{\epsilon}\right), \qquad 0 < \epsilon < \frac{1}{3}. \tag{2.9}$$

Moreover, for $0 < \epsilon \leqslant \frac{2}{3}$, a univariate polynomial exists [44] that maps $[-1,1] \to [-1,1]$, $[-1,-\epsilon] \to [-1,-\frac{2}{3}]$ and $[\epsilon, 1] \to [\frac{2}{3}, 1]$ and has degree $O(1/\epsilon)$. Hence,

$$\deg_{1/3}(f) \leqslant O\left(\frac{1}{\epsilon}\deg_{1-\epsilon}(f)\right), \qquad 0 < \epsilon \leqslant \frac{2}{3}. \tag{2.10}$$

We will need the following dual characterization of the approximate degree.

THEOREM 2.4. *Fix $\epsilon \geqslant 0$. Let $X \subset \mathbb{R}^n$ be a finite set, $f$ a (possibly partial) Boolean function on $X$. Then $\deg_\epsilon(f) > d$ if and only if there exists a function $\psi\colon X \to \mathbb{R}$ such that*

$$\sum_{x \in \operatorname{dom} f} f(x)\psi(x) - \sum_{x \notin \operatorname{dom} f} |\psi(x)| > \epsilon\|\psi\|_1$$

*and* $\sum_X \psi(x)p(x) = 0$ *for every polynomial* $p$ *of degree up to* $d$.

Theorem 2.4 is immediate from linear programming duality; see [48, Sec. 3] for details. In the special case of functions $f\colon X \to \{-1,+1\}$, the first property of $\psi$ simplifies to $\langle f, \psi\rangle > \epsilon\|\psi\|_1$. The following weakening of Theorem 2.4 will be useful in Lemma 3.9.

COROLLARY 2.5. *Fix* $\epsilon \geqslant 0$. *Let* $X \subset \mathbb{R}^n$ *be a finite set,* $f$ *a (possibly partial) partial Boolean function on* $X$. *Then* $\deg_\epsilon(f) > d$ *whenever there exists a function* $\psi\colon X \to \mathbb{R}$ *such that*

$$\sum_{x\in\operatorname{dom} f} f(x)\psi(x) > \frac{1+\epsilon}{2}\|\psi\|_1$$

*and* $\sum_X \psi(x)p(x) = 0$ *for every polynomial* $p$ *of degree up to* $d$.

*Proof.* Substitute $\|\psi\|_1 - \sum_{x\in\operatorname{dom} f} f(x)\psi(x)$ for $\sum_{x\notin\operatorname{dom} f}|\psi(x)|$ in Theorem 2.4. □

The *threshold degree* $\deg_\pm(f)$ of a Boolean function $f\colon X \to \{-1,+1\}$, for a finite subset $X \subset \mathbb{R}^n$, is the limit

$$\deg_\pm(f) = \lim_{\epsilon\searrow 0} \deg_{1-\epsilon}(f). \tag{2.11}$$

Equivalently, $\deg_\pm(f)$ is the least degree of a real polynomial $p$ with $f(x) \equiv \operatorname{sgn} p(x)$.

**2.4. Communication complexity.** For an excellent exposition of quantum communication complexity, see [15], [53]. Here we will mostly limit ourselves to a review of basic facts and notation. Let $f$ be a (possibly partial) Boolean function on the Cartesian product $X \times Y$ of two finite sets $X, Y$. A quantum protocol is said to compute $f$ with error $\epsilon$ if on every input $(x, y) \in \operatorname{dom} f$, the output of the protocol disagrees with the value of $f$ with probability no greater than $\epsilon$. Analogous to classical computation, the cost of a quantum protocol is the maximum number of quantum bits exchanged between the two players on any input $(x, y)$. The least cost of an $\epsilon$-error quantum protocol (with arbitrary prior entanglement) for $f$ is denoted $Q^*_\epsilon(f)$. The precise choice of a constant $\epsilon \in (0, 1/2)$ affects $Q^*_\epsilon(f)$ by at most a constant factor, and thus the setting $\epsilon = 1/3$ entails no loss of generality. As another convention, by the communication complexity of a (possibly partial) sign matrix $F = [F_{ij}]_{i\in I, j\in J}$ will be meant the communication complexity of the associated (possibly partial) Boolean function $f$ on $I \times J$ given by $f(i, j) = F_{ij}$ when $F_{ij} = \pm 1$ and undefined otherwise.

We will additionally consider the setting where the quantum protocol simultaneously solves $n$ communication problems (equivalently, sign matrices) $F_1, F_2, \ldots, F_n$. Given $n$ input instances $(x_1, y_1), \ldots, (x_n, y_n)$, one per communication problem, the protocol is required to output a string $z \in \{-1,+1\}^n$ representing a guess at the vector $(F_1(x_1, y_1), \ldots, F_n(x_n, y_n)) \in \{-1,+1\}^n$. As before, a $(1-\sigma)$-error protocol is one whose output differs from the correct answer with probability no greater than $1-\sigma$, on any given input. The least cost of such a protocol for $F_1, F_2, \ldots, F_n$ is denoted $Q^*_{1-\sigma}(F_1, F_2, \ldots, F_n)$. As usual, we allow $F_1, F_2, \ldots, F_n$ to be partial functions (equivalently, partial sign matrices).

In the case of $n$ communication problems $F_1, F_2, \ldots, F_n$, it is meaningful to consider protocols that solve all but $m$ of the $n$ instances, where the ratio $m/n$ is a small constant. In other words, given $n$ input instances $(x_1, y_1), \ldots, (x_n, y_n)$, one per communication problem, the protocol is required to output, with probability at least $\sigma$, a vector $z \in \{-1, +1\}^n$ such that $z_i = F_i(x_i, y_i)$ for at least $n - m$ indices $i$. We let $Q^*_{1-\sigma,m}(F_1, F_2, \ldots, F_n)$ stand for the least cost of such a quantum protocol for $F_1, F_2, \ldots, F_n$. When referring to this formalism, we will write that a protocol "solves with probability $\sigma$ at least $n - m$ of the communication problems $F_1, F_2, \ldots, F_n$."

Yao [56], Kremer [34], and Razborov [43] showed that the matrix of acceptance probabilities of a low-cost quantum protocol has low trace norm. Using work by Lee, Shraibman, and Špalek [37, Thm. 9], one can show that the $\gamma_2$ norm of a matrix $M$ is the maximum trace norm over all matrices obtained from $M$ by duplicating rows and columns as desired (and normalizing in each case for the number of entries in the resulting matrix). It follows that the matrix of acceptance probabilities of a low-cost quantum protocol has low $\gamma_2$ norm. We will however use an earlier, first-principles proof of this statement, due to Linial and Shraibman [39, Lem. 12], which achieves optimal constants.

THEOREM 2.6 (Linial and Shraibman). *Let $\Pi$ be a quantum protocol of cost $c$, with or without prior entanglement. Then the matrix $M = [\mathbf{P}[\Pi \text{ accepts } (x, y)]]_{x,y}$ satisfies $\gamma_2(M) \leqslant 2^c$.*

This result has the following corollary [39, Thm. 13], described earlier as the generalized discrepancy method.

THEOREM 2.7 (Linial and Shraibman). *Let $F$ be a sign matrix. Then for all $\epsilon \in (0, 1/2)$,*

$$Q^*_\epsilon(F) \geqslant \frac{1}{4} \log \left\{\gamma_{2,\frac{\epsilon}{1-\epsilon}}(F)\right\}.$$

*For all partial sign matrices $F$,*

$$Q^*_\epsilon(F) \geqslant \log \left\{\gamma_{2,\frac{\epsilon}{1-\epsilon}}(F)\right\} - 3.$$

The quantity $\mathrm{GDM}_\epsilon(F)$ from the Introduction refers to the lower bound on $Q^*_\epsilon(F)$ given by Theorem 2.7. Note that the discussion in the Introduction was in terms of a different norm $\mu$ and not $\gamma_2$. This substitution, original to [39], is legitimate because the two norms are within a small multiplicative factor; see [36, Sec. 2.3] for details.

**2.5. Query complexity.** One of the basic models of computation is the decision tree. It represents a deterministic algorithm that computes a given function $f\colon \{-1, +1\}^n \to \{-1, +1\}$ on an unknown input $x \in \{-1, +1\}^n$ by querying a few bits of $x$, in an adaptive manner. A randomized algorithm in this model corresponds to a probability distribution on a family of decision trees. Classical query algorithms have natural quantum counterparts; see [9], [53] for a detailed introduction to this model. Analogous to the classical case, the cost of a quantum query algorithm is the number of queries on the worst-case input. The *$\epsilon$-error quantum query complexity* $T_\epsilon(f)$ of a (possibly partial) Boolean function $f$ on $\{-1, +1\}^n$ is the least cost of a quantum algorithm that computes $f(x)$ correctly with probability at least $1 - \epsilon$, on every input $x \in \operatorname{dom} f$.

Quantum query complexity is closely related to polynomial approximation, as discovered by Beals, Buhrman, Cleve, Mosca, and de Wolf [9, Lem. 4.2].

THEOREM 2.8 (Beals et al.). *Let $A$ be a cost-$T$ quantum query algorithm, with input $x \in \{-1,+1\}^n$ and one-bit output. Then the acceptance probability of $A$ is a real polynomial in $x_1, \ldots, x_n$ of degree at most $2T$. In particular,*

$$T_\epsilon(f) \geqslant \frac{1}{2} \deg_{\frac{\epsilon}{1-\epsilon}}(f) \tag{2.12}$$

*for every (possibly partial) Boolean function $f$ on $\{-1,+1\}^n$.*

Theorem 2.8 states the *polynomial method* for proving lower bounds on quantum query complexity. The relationship (2.12) was used in [9] to prove tight lower bounds for all symmetric functions and has been successfully employed in subsequent research, e.g., [13], [2], [1], [32].

Analogous to quantum communication complexity, we will consider query algorithms that simultaneously solve $n$ problems $f_1, f_2, \ldots, f_n$, where each $f_i$ is a (possibly partial) Boolean function on $\{-1,+1\}^m$. We say that a quantum query algorithm *solves with probability $\sigma$ at least $n-m$ of the problems* $f_1, f_2, \ldots, f_n$ if on every input $(x_1, x_2, \ldots, x_n) \in \prod \operatorname{dom} f_i$, the algorithm outputs, with probability at least $\sigma$, a string $z \in \{-1,+1\}^n$ with $z_i = f_i(x_i)$ for $n-m$ or more indices $i$. We let $T_{1-\sigma,m}(f_1, f_2, \ldots, f_n)$ denote the least cost of such a quantum algorithm.

**2.6. Fourier transform.** Consider the vector space of functions $\{-1,+1\}^n \to \mathbb{R}$. For $S \subseteq \{1,2,\ldots,n\}$, define $\chi_S \colon \{-1,+1\}^n \to \{-1,+1\}$ by $\chi_S(x) = \prod_{i \in S} x_i$. Then the functions $\chi_S$, $S \subseteq \{1,2,\ldots,n\}$, form an orthogonal basis for the vector space in question. In particular, every function $f \colon \{-1,+1\}^n \to \mathbb{R}$ has a unique representation of the form

$$f = \sum_{S \subseteq \{1,2,\ldots,n\}} \hat{f}(S)\, \chi_S,$$

where $\hat{f}(S) = 2^{-n} \sum_{x \in \{-1,+1\}^n} f(x)\chi_S(x)$ is the *Fourier coefficient* of $f$ that corresponds to $\chi_S$. The orthogonality of $\{\chi_S\}$ leads to

$$\sum_{S \subseteq \{1,2,\ldots,n\}} \hat{f}(S)^2 = \mathop{\mathbf{E}}_{x \in \{-1,+1\}^n} [f(x)^2], \tag{2.13}$$

a fact known as *Parseval's identity*. The functional

$$\|\hat{f}\|_1 = \sum_{S \subseteq \{1,2,\ldots,n\}} |\hat{f}(S)|$$

is clearly subadditive and submultiplicative. In other words,

$$\|\widehat{f+g}\|_1 \leqslant \|\hat{f}\|_1 + \|\hat{g}\|_1, \tag{2.14}$$

$$\|\widehat{f \circ g}\|_1 \leqslant \|\hat{f}\|_1 \, \|\hat{g}\|_1 \tag{2.15}$$

for all $f, g \colon \{-1,+1\}^n \to \mathbb{R}$.

## 3. Preparatory work

Theorems 2.6 and 2.8 point to a key similarity between quantum communication and query complexity. Specifically, every efficient communication protocol, when viewed as a matrix of acceptance probabilities, resides in the *convex set* corresponding to matrices of low $\gamma_2$ norm. Every efficient query algorithm, when viewed as a function of acceptance probabilities, resides in the *convex set* corresponding to low-degree polynomials. In this section, we develop a number of auxiliary results that are not affected by the nature of the convex set and are thus common to quantum communication and query complexity. This allows us to avoid a considerable duplication of effort. In what follows, we categorize the preparatory work into results pertaining to XOR lemmas, direct product theorems, and direct sum theorems.

**3.1. Auxiliaries for XOR lemmas.** A starting fact in our analysis is a polynomial construction. It will subsequently play a key role in finding the witness object described in the Introduction.

LEMMA 3.1. *For any* $\eta_1, \eta_2, \ldots, \eta_n \in [0,1)$, *define* $\mu = \Pi(\eta_1, \eta_2, \ldots, \eta_n)$ *and* $\eta = \max\{\eta_1, \eta_2, \ldots, \eta_n\}$. *For* $k = 0, 1, 2, \ldots, n-1$, *let* $p_k \colon [-1,1]^n \to \mathbb{R}$ *be the unique degree-$k$ multilinear polynomial such that*

$$p_k(z) = (-1)^k \prod_{i=1}^{k} (|z| - i), \qquad z \in \{-1,+1\}^n.$$

*Then*

$$\mathop{\mathbf{E}}_{\mu}[|p_k(z)|] \leqslant p_k(1^n)\mu(1^n) \left\{ 1 + \binom{n}{k+1} \frac{\eta^{k+1}}{(1-\eta)^n} \right\}, \tag{3.1}$$

$$\|\hat{p}_k\|_1 \leqslant k! \binom{n+k}{k}. \tag{3.2}$$

*Furthermore,* $p_k(z) \geqslant 0$ *for all* $z \in [-1,1]^n$ *provided that $k$ is even.*

*Proof.* Nonnegativity for even $k$ is immediate on $\{-1,+1\}^n$ and generalizes to all of $[-1,1]^n$ by the multilinearity of $p_k$ and convexity. Next, (2.14) and (2.15) imply

$$\|\hat{p}_k\|_1 \leqslant \prod_{i=1}^{k} (n+i) = k! \binom{n+k}{k}.$$

It remains to prove (3.1). We have

$$\left.\begin{aligned} &p_k(1^n) = k!, \\ &p_k(z) = 0, && 1 \leqslant |z| \leqslant k, \\ &|p_k(z)| = p_k(1^n) \binom{|z|-1}{k} \\ &\qquad \leqslant p_k(1^n) \binom{n}{k+1} \binom{n-k-1}{|z|-k-1} \binom{n}{|z|}^{-1}, && |z| \geqslant k+1. \end{aligned}\right\} \tag{3.3}$$

In addition,

$$\mathop{\mathbf{P}}_{\mu}[|z| = i] = \sum_{|S|=i} \prod_{j\in S} \eta_j \cdot \prod_{j\notin S}(1-\eta_j) = \mu(1^n) \sum_{|S|=i} \prod_{j\in S} \frac{\eta_j}{1-\eta_j}$$
$$\leqslant \mu(1^n)\binom{n}{i}\left(\frac{\eta}{1-\eta}\right)^i, \tag{3.4}$$

where the final step uses $\eta_j \leqslant \eta$. Putting together (3.3) and (3.4) gives:

$$\mathop{\mathbf{E}}_{\mu}[|p_k(z)|] \leqslant p_k(1^n)\mu(1^n) + \sum_{i=k+1}^{n} p_k(1^n)\binom{n}{k+1}\binom{n-k-1}{i-k-1}\mu(1^n)\left(\frac{\eta}{1-\eta}\right)^i$$
$$= p_k(1^n)\mu(1^n)\left\{1 + \binom{n}{k+1}\left(\frac{\eta}{1-\eta}\right)^{k+1} \sum_{i=0}^{n-k-1}\binom{n-k-1}{i}\left(\frac{\eta}{1-\eta}\right)^i\right\}$$
$$= p_k(1^n)\mu(1^n)\left\{1 + \binom{n}{k+1}\frac{\eta^{k+1}}{(1-\eta)^n}\right\}. \qquad \square$$

Using the polynomial $p_k$ from the previous lemma, we will now construct the desired witness object $\Psi_k$, for later use in the XOR lemmas. For now we will establish only those properties of $\Psi_k$ that are common to the settings of communication and query complexity.

LEMMA 3.2. *Fix $\epsilon \in (0,1)$. Consider a (possibly partial) Boolean function $g_i$ on a finite set $X_i$ $(i = 1, 2, \dots, n)$. Let $\psi_i\colon X_i \to \mathbb{R}$ be given with*

$$\|\psi_i\|_1 = 1, \qquad i = 1, 2, \dots, n, \tag{3.5}$$
$$\sum_{x\in \operatorname{dom} g_i} g_i(x_i)\psi_i(x_i) - \sum_{x\notin \operatorname{dom} g_i} |\psi_i(x_i)| > (1-\epsilon)\|\psi_i\|_1, \qquad i = 1, 2, \dots, n. \tag{3.6}$$

*For each $i$, let $f_i\colon X_i \to \{-1,+1\}$ be the extension of $g_i$ given by $f_i(x_i) = -\widetilde{\operatorname{sgn}}\, \psi_i(x_i)$ outside* $\operatorname{dom} g_i$. *For $k = 0, 1, 2, \dots, n-1$, define $\Psi_k\colon \prod X_i \to \mathbb{R}$ by*

$$\Psi_k(x_1, \dots, x_n) = p_k(\dots, f_i(x_i)\operatorname{sgn}\psi_i(x_i), \dots) \prod_{i=1}^{n} \psi_i(x_i),$$

*where $p_k$ is the degree-$k$ polynomial from Lemma* 3.1. *Then for all $\delta \geqslant 0$,*

$$\sum_{x\in\prod \operatorname{dom} g_i} \Psi_k(x)\prod_{i=1}^{n} g_i(x_i) - \sum_{x\notin\prod \operatorname{dom} g_i} |\Psi_k(x)| - \delta\|\Psi_k\|_1$$
$$> k!\left(1-\frac{\epsilon}{2}\right)^n\left\{1 - \delta - (1+\delta)\binom{n}{k+1}\frac{\left(\frac{1}{2}\epsilon\right)^{k+1}}{\left(1-\frac{1}{2}\epsilon\right)^n}\right\}. \tag{3.7}$$

*Proof.* It is clear that $\langle f_i, \psi_i\rangle > (1-\epsilon)\|\psi_i\|_1 = 1-\epsilon$ for each $i$. Let $\eta_i = \frac{1}{2} - \frac{1}{2}\langle f_i, \psi_i\rangle$ and $\eta = \max\{\eta_1, \eta_2, \dots, \eta_n\}$. Then

$$\eta < \frac{\epsilon}{2}. \tag{3.8}$$

Let $\lambda$ be the probability distribution on $\prod X_i$ given by $\lambda(\ldots, x_i, \ldots) = \prod |\psi_i(x_i)|$. When $(\ldots, x_i, \ldots) \sim \lambda$, the string $(\ldots, f_i(x_i) \operatorname{sgn} \psi_i(x_i), \ldots) \in \{-1,+1\}^n$ is distributed according to $\mu = \Pi(\eta_1, \eta_2, \ldots, \eta_n)$. As a result,

$$\begin{aligned} \|\Psi_k\|_1 &= \sum_{X_1 \times \cdots \times X_n} |p_k(\ldots, f_i(x_i) \operatorname{sgn} \psi_i(x_i), \ldots)| \prod_{i=1}^n |\psi_i(x_i)| \\ &= \mathop{\mathbf{E}}_{x \sim \lambda} [|p_k(\ldots, f_i(x_i) \operatorname{sgn} \psi_i(x_i), \ldots)|] \\ &= \mathop{\mathbf{E}}_{z \sim \mu} [|p_k(z)|]. \end{aligned} \tag{3.9}$$

Let $D \subseteq \prod \operatorname{dom} g_i$ be given by $D = \prod \{x_i \in X_i : f_i(x_i) = \operatorname{sgn} \psi_i(x_i)\}$. Then

$$\begin{aligned} &\sum_{x \in \prod \operatorname{dom} g_i} \Psi_k(x) \prod_{i=1}^n g_i(x_i) - \sum_{x \notin \prod \operatorname{dom} g_i} |\Psi_k(x)| - \delta \|\Psi_k\|_1 \\ &\qquad \geqslant \sum_{x \in D} \Psi_k(x) \prod_{i=1}^n f_i(x_i) - \sum_{x \notin D} |\Psi_k(x)| - \delta \|\Psi_k\|_1 \\ &\qquad \geqslant 2 \sum_{x \in D} \Psi_k(x) \prod_{i=1}^n f_i(x_i) - (1+\delta) \|\Psi_k\|_1 \\ &\qquad = 2\mu(1^n) p_k(1^n) - (1+\delta) \|\Psi_k\|_1 \\ &\qquad = 2\mu(1^n) p_k(1^n) - (1+\delta) \mathop{\mathbf{E}}_{z \sim \mu} [|p_k(z)|] && \text{by (3.9)} \\ &\qquad \geqslant \mu(1^n) p_k(1^n) \left\{ 2 - (1+\delta) \left( 1 + \binom{n}{k+1} \frac{\eta^{k+1}}{(1-\eta)^n} \right) \right\} && \text{by (3.1).} \end{aligned}$$

In view of (3.8) and the bound $\mu(1^n) p_k(1^n) \geqslant k!(1-\eta)^n > k!(1-\epsilon/2)^n$, the proof is complete. ∎

A common operation in this manuscript is bounding the correlation of a given function with the elements of a given convex set. In the case of quantum query complexity, this operation is effortless because of the way polynomial multiplication is defined. More care is needed in the setting of quantum communication complexity, where this step corresponds to bounding the norm dual to the convex set. We address the latter case below.

LEMMA 3.3. *Fix finite sets $X_1, X_2, \ldots, X_n$. Let $\mathbb{R}^{X_1}, \mathbb{R}^{X_2}, \ldots, \mathbb{R}^{X_n}$, and $\mathbb{R}^{\prod X_i}$ be normed by $|||\cdot|||$, where*

$$C_1 = \max \left\{ \frac{|||\bigotimes_{i=1}^n \phi_i|||^*}{\prod_{i=1}^n |||\phi_i|||^*} : \phi_i \in \mathbb{R}^{X_i} \setminus \{0\} \text{ for all } i = 1, 2, \ldots, n \right\}, \tag{3.10}$$

$$C_2 = \max \left\{ \frac{\|\phi\|_\infty}{|||\phi|||} : \phi \in \mathbb{R}^{X_i} \setminus \{0\} \text{ for some } i = 1, 2, \ldots, n \right\}. \tag{3.11}$$

*For $k = 0, 1, 2, \ldots, n$, let $\mathscr{C}_k$ be the convex hull of functions $\xi \colon \prod X_i \to \mathbb{R}$ of the form*

$$\xi(x_1, \ldots, x_n) = \mathop{\mathbf{E}}_{|S|=k} \left[ \prod_{i \in S} \xi_{S,i}(x_i) \right], \tag{3.12}$$

*where each* $\xi_{S,i}\colon X_i \to \mathbb{R}$ *obeys* $\|\xi_{S,i}\|_\infty \leqslant 1$. *Then for all* $\psi_i\colon X_i \to \mathbb{R}$ *with* $\|\psi_i\|_1 \leqslant 1$ $(i = 1, 2, \ldots, n)$,

$$\max_{\zeta \in \mathscr{C}_k} \left|\!\left|\!\left| \zeta \circ \bigotimes_{i=1}^n \psi_i \right|\!\right|\!\right|^* \leqslant C_1 C_2^k \mathop{\mathbf{E}}_{|S|=n-k} \left[ \prod_{i \in S} |||\psi_i|||^* \right].$$

*Proof.* By convexity, it suffices to prove the claim for the functions $\xi$ in (3.12). In view of (3.10),

$$\left|\!\left|\!\left| \xi \circ \bigotimes_{i=1}^n \psi_i \right|\!\right|\!\right|^* \leqslant C_1 \mathop{\mathbf{E}}_{|S|=k} \left[ \prod_{i \in S} |||\xi_{S,i} \circ \psi_i|||^* \cdot \prod_{i \notin S} |||\psi_i|||^* \right].$$

By duality, (3.11) is equivalent to saying that $|||\phi|||^* \leqslant C_2 \|\phi\|_1$ for all $\phi \in \bigcup \mathbb{R}^{X_i}$, whence

$$\begin{aligned} \left|\!\left|\!\left| \xi \circ \bigotimes_{i=1}^n \psi_i \right|\!\right|\!\right|^* &\leqslant C_1 C_2^k \mathop{\mathbf{E}}_{|S|=k} \left[ \prod_{i \in S} \|\xi_{S,i} \circ \psi_i\|_1 \cdot \prod_{i \notin S} |||\psi_i|||^* \right] \\ &\leqslant C_1 C_2^k \mathop{\mathbf{E}}_{|S|=n-k} \left[ \prod_{i \in S} |||\psi_i|||^* \right]. \end{aligned}$$

□

**3.2. Auxiliaries for direct product theorems.** We now turn our attention to the setting of direct product theorems. We start with a relaxed formalization of what it means to simultaneously solve $n$ problems.

DEFINITION 3.4 (Approximants). Fix a (possibly partial) Boolean function $g_i$ on a finite set $X_i$, $i = 1, 2, \ldots, n$. A $(\sigma, m)$-*approximant* for $(g_1, g_2, \ldots, g_n)$ is any system $\{\phi_z\}$ of functions $\phi_z\colon \prod X_i \to \mathbb{R}$, $z \in \{-1, +1\}^n$, such that:

$$\sum_{z \in \{-1,+1\}^n} |\phi_z(x_1, \ldots, x_n)| \leqslant 1, \qquad x \in \prod X_i, \tag{3.13}$$

$$\sum_{|z| \leqslant m} \phi_{(z_1 g_1(x_1), \ldots, z_n g_n(x_n))}(x_1, \ldots, x_n) \geqslant \sigma, \qquad x \in \prod \operatorname{dom} g_i. \tag{3.14}$$

It is straightforward to see, as we will in sections to come, that communication protocols and query algorithms that solve with probability $\sigma$ at least $n - m$ of the problems $g_1, g_2, \ldots, g_n$ give rise to representations $\{\phi_z\}$ that obey (3.13) and (3.14). The representations $\{\phi_z\}$ that arise in that way will obey various additional properties, but we will only appeal to (3.13) and (3.14) in the proofs of our lower bounds. As the reader may have already guessed, strong direct product theorems correspond to $m = 0$, whereas threshold direct product theorems correspond to $m = \beta n$ for some small constant $\beta > 0$.

We now recall a result on the polynomial approximation of symmetric functions due to de Wolf [54], improving on earlier work in [47]. We only require a rather special case of de Wolf's theorem.

THEOREM 3.5 (De Wolf). *Let $\alpha > 0$ be a sufficiently small absolute constant. Then for all integers $m, \ell \geqslant 0$, there is a degree-$\ell$ univariate polynomial $Q_\ell$ with*

$$|Q_\ell(i) - (-1)^i| \leqslant 2^{-\frac{\alpha\ell^2}{n}+m+1}, \qquad i = 0, 1, \dots, m, \tag{3.15}$$

$$|Q_\ell(i)| \leqslant 2^{-\frac{\alpha\ell^2}{n}+m+1}, \qquad i = m+1, m+2, \dots, n, \tag{3.16}$$

$$|Q_\ell(i)| \leqslant 1, \qquad i = 0, 1, \dots, n.$$

In words, Theorem 3.5 gives a polynomial of reasonably low degree that approximates the parity function with extremely high accuracy at the integer points in $[0, m]$ and is exponentially close to zero at the integer points in $(m, n]$. We will need the following corollary to Theorem 3.5.

COROLLARY 3.6. *Let $\alpha$ be the absolute constant from Theorem 3.5. Then for all integers $m, \ell \geqslant 0$, there is a degree-$\ell$ symmetric polynomial $q_\ell\colon \{-1,+1\}^n \to [-1,1]$ such that:*

$$\left| q_\ell(z) - \prod_{i=1}^n z_i \right| \leqslant 2^{-\frac{\alpha\ell^2}{n}+m+1}, \qquad |z| \leqslant m, \tag{3.17}$$

$$|q_\ell(z)| \leqslant 2^{-\frac{\alpha\ell^2}{n}+m+1}, \qquad |z| > m, \tag{3.18}$$

$$\|\hat{q}_\ell\|_1 \leqslant \binom{n}{\leqslant \ell}^{1/2}. \tag{3.19}$$

*Proof.* Put $q_\ell(z) = Q_\ell(|z|)$, where $Q_\ell$ is the polynomial from Theorem 3.5. Then (3.17) and (3.18) follow at once from the analogous properties of $Q_\ell$. Since $q_\ell$ sends $\{-1,+1\}^n \to [-1,1]$, one infers (3.19) from Parseval's identity (2.13):

$$\sum_{|S|\leqslant \ell} |\hat{q}_\ell(S)| \leqslant \left( \sum_{|S|\leqslant\ell} \hat{q}_\ell(S)^2 \right)^{\frac{1}{2}} \binom{n}{\leqslant \ell}^{\frac{1}{2}} = \left( \mathop{\mathbf{E}}_{z\in\{-1,+1\}^n} [q_\ell(z)^2] \right)^{\frac{1}{2}} \binom{n}{\leqslant \ell}^{\frac{1}{2}}. \qquad \Box$$

We are now prepared to prove the key technical lemma that will allow us to obtain direct product theorems for communication and query complexity.

LEMMA 3.7. *Consider a (possibly partial) Boolean function $g_i$ on a finite set $X_i$, for $i = 1, 2, \dots, n$. Let $\psi_i\colon X_i \to \mathbb{R}$ be given that obeys (3.5) and (3.6). Define $f_i\colon X_i \to \{-1,+1\}$ $(i = 1, 2, \dots, n)$ and $\Psi_k\colon \prod X_i \to \mathbb{R}$ $(k = 0, 1, 2, \dots, n-1)$ as in Lemma 3.2. For a given $(\sigma, m)$-approximant $\{\phi_z\}$ of $(g_1, g_2, \dots, g_n)$, let $\Phi_\ell\colon \prod X_i \to \mathbb{R}$ be defined by*

$$\Phi_\ell(x_1, \dots, x_n) = \sum_{z\in\{-1,+1\}^n} \phi_z(x_1, \dots, x_n) q_\ell(\dots, z_i f_i(x_i), \dots) \prod_{i=1}^n z_i,$$

*where $q_\ell$ is the degree-$\ell$ polynomial from Corollary 3.6. Then*

$$\langle \Phi_\ell, \Psi_k \rangle$$
$$> k!\left(1 - \frac{\epsilon}{2}\right)^n \left\{ 2 - \left(2 - \sigma + 2^{-\frac{\alpha\ell^2}{n}+m+1}\right)\left(1 + \binom{n}{k+1} \frac{\left(\frac{1}{2}\epsilon\right)^{k+1}}{\left(1-\frac{1}{2}\epsilon\right)^n}\right)\right\}, \tag{3.20}$$

*where $\alpha > 0$ is the absolute constant from Theorem* 3.5.

*Proof.* In view of (3.13) and the fact that $q_\ell$ sends $\{-1,+1\}^n \to [-1,1]$, we have

$$\|\Phi_\ell\|_\infty \leqslant 1. \tag{3.21}$$

Now,

$$\begin{aligned}
\langle \Phi_\ell, \Psi_k \rangle &= \sum_{x \notin \prod \operatorname{dom} g_i} \Phi_\ell(x)\Psi_k(x) \;+ \sum_{x \in \prod \operatorname{dom} g_i} \Psi_k(x) \prod_{i=1}^n g_i(x_i) \\
&\qquad + \sum_{x \in \prod \operatorname{dom} g_i} \Psi_k(x) \left\{ \Phi_\ell(x) - \prod_{i=1}^n g_i(x_i) \right\} \\
&\geqslant \sum_{x \in \prod \operatorname{dom} g_i} \Psi_k(x) \prod_{i=1}^n g_i(x_i) \;- \sum_{x \notin \prod \operatorname{dom} g_i} |\Psi_k(x)| \\
&\qquad - \|\Psi_k\|_1 \max_{x \in \prod \operatorname{dom} g_i} \left| \Phi_\ell(x) - \prod_{i=1}^n g_i(x_i) \right|,
\end{aligned}$$

where the second step uses (3.21). In view of (3.7), it remains to prove that

$$\left| \Phi_\ell(x) - \prod_{i=1}^n g_i(x) \right| \leqslant 1 - \sigma + 2^{-\frac{\alpha \ell^2}{n} + m + 1}, \qquad x \in \prod \operatorname{dom} g_i. \tag{3.22}$$

For this, define $G\colon \{-1,+1\}^n \to \{-1,0,+1\}$ by

$$G(z) = \begin{cases} z_1 z_2 \cdots z_n, & |z| \leqslant m, \\ 0, & \text{otherwise.} \end{cases}$$

For every $x = (x_1, \ldots, x_n) \in \prod \operatorname{dom} g_i$,

$$\begin{aligned}
&\left| \Phi_\ell(x) - \prod_{i=1}^n g_i(x_i) \right| \\
&\qquad = \left| \Phi_\ell(x) \prod_{i=1}^n g_i(x_i) - 1 \right| \\
&\qquad = \left| \sum_{z \in \{-1,+1\}^n} \phi_z(x)\, q_\ell(\ldots, z_i g_i(x_i), \ldots) \prod_{i=1}^n z_i g_i(x_i) - 1 \right| \\
&\qquad \leqslant \left| \sum_{z \in \{-1,+1\}^n} \phi_z(x)\, G(\ldots, z_i g_i(x_i), \ldots) \prod_{i=1}^n z_i g_i(x_i) - 1 \right| \\
&\qquad\qquad + \sum_{z \in \{-1,+1\}^n} |\phi_z(x)|\, \|G - q_\ell\|_\infty \\
&\qquad = \underbrace{\left| \sum_{|z| \leqslant m} \phi_{(z_1 g_1(x_1), \ldots, z_n g_n(x_n))}(x) \;-\; 1 \right|}_{\leqslant 1 - \sigma} + \|G - q_\ell\|_\infty \underbrace{\sum_{z \in \{-1,+1\}^n} |\phi_z(x)|}_{\leqslant 1},
\end{aligned}$$

where the indicated bounds in the final step follow by (3.13) and (3.14). Corollary 3.6 guarantees that $\|G - q_\ell\|_\infty \leqslant 2^{-\alpha\ell^2/n+m+1}$, which settles (3.22) and the lemma. ∎

**3.3. Auxiliaries for direct sum theorems.** When the given problems $f_1, f_2, \ldots, f_n$ are of comparable complexity, it makes sense to speak of XOR lemmas and direct product theorems. When the relative complexities of the problems vary greatly, one can only hope to prove a direct sum theorem. We develop the needed technical tools below, for communication and query complexity.

LEMMA 3.8. *Fix finite sets $X_1, X_2, \ldots, X_n$. Let $\mathbb{R}^{X_1}, \mathbb{R}^{X_2}, \ldots, \mathbb{R}^{X_n}$, and $\mathbb{R}^{\prod X_i}$ be normed by $|\!|\!|\cdot|\!|\!|$, with $C_1$ defined by* (3.10). *Then for all $\epsilon_1, \epsilon_2, \ldots, \epsilon_n \in (0,1)$ and all functions $f_i\colon X_i \to \{-1,+1\}$ $(i = 1,2,\ldots,n)$,*

$$\left|\!\left|\!\left| \bigotimes_{i=1}^n f_i \right|\!\right|\!\right|_{\prod \epsilon_i} \geqslant \frac{1}{C_1} \prod_{i=1}^n |\!|\!| f_i |\!|\!|_{\epsilon_i}. \tag{3.23}$$

*For all* (*possibly partial*) *Boolean functions $g_i$ on $X_i$ $(i = 1,2,\ldots,n)$,*

$$\left|\!\left|\!\left| \bigotimes_{i=1}^n g_i \right|\!\right|\!\right|_{2\prod \epsilon_i - 1} \geqslant \frac{2}{C_1} \prod_{i=1}^n |\!|\!| g_i |\!|\!|_{\epsilon_i}. \tag{3.24}$$

*Proof.* By Corollary 2.2, we can fix $\psi_i\colon X_i \to \mathbb{R}$ such that

$$|\!|\!| f_i |\!|\!|_{\epsilon_i} = \frac{\langle f_i, \psi_i\rangle - \epsilon_i \|\psi_i\|_1}{|\!|\!|\psi_i|\!|\!|^*}, \qquad i = 1,2,\ldots,n. \tag{3.25}$$

Then

$$\begin{aligned}
\left|\!\left|\!\left| \bigotimes_{i=1}^n f_i \right|\!\right|\!\right|_{\prod \epsilon_i} &\geqslant \frac{\langle \bigotimes f_i, \bigotimes \psi_i\rangle - \|\bigotimes \psi_i\|_1 \prod \epsilon_i}{|\!|\!|\bigotimes \psi_i|\!|\!|^*} && \text{by Corollary 2.2} \\
&\geqslant \frac{\prod \langle f_i, \psi_i\rangle - \prod \epsilon_i \|\psi_i\|_1}{C_1 \prod |\!|\!|\psi_i|\!|\!|^*} && \text{by (3.10)} \\
&= \frac{\prod \langle f_i, \psi_i\rangle - \prod \epsilon_i \|\psi_i\|_1}{C_1 \prod (\langle f_i, \psi_i\rangle - \epsilon_i \|\psi_i\|_1)} \prod |\!|\!| f_i |\!|\!|_{\epsilon_i} && \text{by (3.25)} \\
&\geqslant \frac{1}{C_1} \prod |\!|\!| f_i |\!|\!|_{\epsilon_i},
\end{aligned}$$

where the final step follows from the fact that $\prod A_i - \prod a_i \geqslant \prod (A_i - a_i)$ for any reals $A_1, \ldots, A_n, a_1, \ldots, a_n$ with $A_i \geqslant a_i \geqslant 0$, $i = 1,2,\ldots,n$. This completes the proof of (3.23).

The proof of (3.24) is closely analogous. Corollary 2.2 provides $\psi_i\colon X_i \to \mathbb{R}$ with

$$|\!|\!| g_i |\!|\!|_{\epsilon_i} \leqslant \frac{1}{|\!|\!|\psi_i|\!|\!|^*} \left\{ \sum_{x_i \in \operatorname{dom} g_i} g_i(x_i)\psi_i(x_i) - \epsilon_i \|\psi_i\|_1 \right\}, \qquad i = 1,2,\ldots,n. \tag{3.26}$$

Then

$$\begin{aligned}
&\left\|\bigotimes_{i=1}^{n} g_i\right\|_{2\prod\epsilon_i-1} \\
&\quad\geqslant \frac{2}{\|\bigotimes\psi_i\|^*}\left\{\sum_{\prod \operatorname{dom} g_i}\prod_i g_i(x_i)\psi_i(x_i) - \|\bigotimes\psi_i\|_1\prod_i\epsilon_i\right\} && \text{by Corollary 2.2}\\
&\quad\geqslant \frac{2}{C_1\prod\|\psi_i\|^*}\left\{\sum_{\prod \operatorname{dom} g_i}\prod_i g_i(x_i)\psi_i(x_i) - \|\bigotimes\psi_i\|_1\prod_i\epsilon_i\right\} && \text{by (3.10)}\\
&\quad\geqslant \frac{\prod_i\left\{\sum_{\operatorname{dom} g_i} g_i(x_i)\psi_i(x_i)\right\} - \prod_i \epsilon_i\|\psi_i\|_1}{C_1\prod_i\left\{\sum_{\operatorname{dom} g_i} g_i(x_i)\psi_i(x_i) - \epsilon_i\|\psi_i\|_1\right\}}\cdot 2\prod\|g_i\|_{\epsilon_i} && \text{by (3.26)}\\
&\quad\geqslant \frac{2}{C_1}\prod\|g_i\|_{\epsilon_i},
\end{aligned}$$

where the final step follows from the fact that $\prod A_i - \prod a_i \geqslant \prod(A_i - a_i)$ for any reals $A_1,\dots,A_n,a_1,\dots,a_n$ with $A_i \geqslant a_i \geqslant 0$, $i = 1,2,\dots,n$. This settles (3.24). ∎

An analogous result holds for polynomial approximation:

LEMMA 3.9. *Fix finite sets $X_1, X_2, \dots, X_n \subset \mathbb{R}^m$. Then for all $\epsilon_1,\epsilon_2,\dots,\epsilon_n \in (0,1)$ and all functions $f_i\colon X_i \to \{-1,+1\}$ $(i = 1,2,\dots,n)$,*

$$\deg_{\prod\epsilon_i}\left(\bigotimes_{i=1}^{n} f_i\right) \geqslant \sum_{i=1}^{n}\deg_{\epsilon_i}(f_i). \tag{3.27}$$

*For all (possibly partial) Boolean functions $g_i$ on $X_i$ $(i = 1,2,\dots,n)$,*

$$\deg_{2\prod\epsilon_i-1}\left(\bigotimes_{i=1}^{n} g_i\right) \geqslant \sum_{i=1}^{n}\deg_{\epsilon_i}(g_i). \tag{3.28}$$

*Proof.* We may clearly assume that each $f_i$ is nonconstant. By Theorem 2.4, for each $i$ there exists $\psi_i\colon X_i \to \mathbb{R}$ with

$$\langle f_i, \psi_i\rangle > \epsilon_i\|\psi_i\|_1$$

and $\sum_{x_i\in X_i}\psi_i(x_i)p(x_i) = 0$ for every polynomial $p$ of degree less than $\deg_{\epsilon_i}(f_i)$. Then clearly $\langle\bigotimes f_i, \bigotimes\psi_i\rangle > \|\bigotimes\psi_i\|_1\prod\epsilon_i$ and $\sum_{\prod X_i} p(\dots,x_i,\dots)\prod\psi_i(x_i) = 0$ for every polynomial $p$ of degree less than $\sum\deg_{\epsilon_i}(f_i)$. Again by Theorem 2.4, the proof of (3.27) is complete.

The proof of (3.28) is similar. We may assume that each $g_i$ is nonconstant. By Theorem 2.4, for each $i$ there exists $\psi_i\colon X_i \to \mathbb{R}$ with

$$\sum_{x_i\in\operatorname{dom} g_i} g_i(x_i)\psi_i(x_i) > \epsilon_i\|\psi_i\|_1$$

and $\sum_{x_i \in X_i} \psi_i(x_i)p(x_i) = 0$ for every polynomial $p$ of degree less than $\deg_{\epsilon_i}(g_i)$. Letting $\psi = \bigotimes \psi_i$, we have

$$\sum_{x \in \prod \operatorname{dom} g_i} \psi(x) \prod_{i=1}^{n} g_i(x_i) > \|\psi\|_1 \prod_{i=1}^{n} \epsilon_i$$

and $\sum_{\prod X_i} p(\dots, x_i, \dots) \prod \psi_i(x_i) = 0$ for every polynomial $p$ of degree less than $\sum \deg_{\epsilon_i}(f_i)$. By Corollary 2.5, the proof of (3.28) is complete. ∎

## 4. Quantum communication

This section is devoted to our results on quantum communication complexity. In Section **4.1**, we prove XOR lemmas and direct product theorems for any approximate norm whose dual exhibits submultiplicative behavior. In the subsections that follow, we specialize our results to $\gamma_2$, obtaining XOR lemmas, direct product theorems, and direct sum theorems for communication complexity.

**4.1. Solution for arbitrary norms.** In what follows, $|||\cdot|||$ stands for any norm on Euclidean space. The results below are meaningful as long as the dual norm behaves nicely under tensor product, viz., a reasonable bound can be placed on $|||\bigotimes \psi_i|||^*$ in terms of $\prod |||\psi_i|||^*$. We start with an XOR lemma.

THEOREM 4.1. *Fix finite sets $X_1, X_2, \dots, X_n$. Let $\mathbb{R}^{X_1}, \mathbb{R}^{X_2}, \dots, \mathbb{R}^{X_n}$, and $\mathbb{R}^{\prod X_i}$ be normed by $|||\cdot|||$, with $C_1, C_2$ defined by* (3.10) *and* (3.11)*. Fix a* (*possibly partial*) *Boolean function $g_i$ on $X_i$ $(i = 1, 2, \dots, n)$. Then for every $\epsilon, \delta \in (0,1)$ and $k = 0, 1, 2, \dots, n-1$,*

$$\left|\left|\left|\bigotimes_{i=1}^{n} g_i\right|\right|\right|_\delta \geqslant \frac{\displaystyle\prod_{i=1}^{n} |||g_i|||_{1-\epsilon}}{\displaystyle\mathop{\mathbf{E}}_{|S|=k}\left[\prod_{i\in S} |||g_i|||_{1-\epsilon}\right]} \cdot \frac{1 - \delta - (1+\delta)\dbinom{n}{k+1}\dfrac{\left(\frac{1}{2}\epsilon\right)^{k+1}}{\left(1-\frac{1}{2}\epsilon\right)^{n}}}{\dfrac{\epsilon^{n-k}}{\left(1-\frac{1}{2}\epsilon\right)^{n}}\dbinom{n+k}{k} C_1 C_2^k}.$$

*Proof.* By Corollary 2.2, for each $i$ there exists $\psi_i \colon X_i \to \mathbb{R}$ such that

$$|||g_i|||_{1-\epsilon} = \frac{1}{|||\psi_i|||^*}\left\{\sum_{x_i \in \operatorname{dom} g_i} g_i(x_i)\psi_i(x_i) - \sum_{x_i \notin \operatorname{dom} g_i} |\psi_i(x_i)| - (1-\epsilon)\|\psi_i\|_1\right\}. \quad (4.1)$$

In particular, the expression in braces is positive for all $i$. By homogeneity, we may assume that $\|\psi_i\|_1 = 1$ for all $i$. Define $\Psi_k \colon \prod X_i \to \mathbb{R}$ as in Lemma 3.2. By Corollary 2.2,

$$\left|\left|\left|\bigotimes_{i=1}^{n} g_i\right|\right|\right|_\delta \geqslant \frac{1}{|||\Psi_k|||^*}\left\{\sum_{x \in \prod \operatorname{dom} g_i} \Psi_k(x) \prod_{i=1}^{n} g_i(x_i) - \sum_{x \notin \prod \operatorname{dom} g_i} |\Psi_k(x)| - \delta\|\Psi_k\|_1\right\}.$$

In view of (3.7), it remains to prove that

$$|||\Psi_k|||^* \leqslant k!\binom{n+k}{k}\epsilon^{n-k} C_1 C_2^k \mathop{\mathbf{E}}_{|S|=n-k}\left[\prod_{i\in S} \frac{1}{|||g_i|||_{1-\epsilon}}\right]. \quad (4.2)$$

For this, note first that (4.1) gives

$$|||\psi_i|||^* \leqslant \frac{\epsilon\|\psi_i\|_1}{|||g_i|||_{1-\epsilon}} = \frac{\epsilon}{|||g_i|||_{1-\epsilon}}, \qquad i = 1, 2, \dots, n. \tag{4.3}$$

Let $\mathscr{C}_0, \mathscr{C}_1, \dots, \mathscr{C}_n$ be as defined in Lemma 3.3. In light of (3.2) and the symmetry of the polynomial $p_k$ from Lemma 3.1, one has

$$p_k(\dots, f_i(x_i) \operatorname{sgn} \psi_i(x_i), \dots) \in k!\binom{n+k}{k}\mathscr{C}_k, \tag{4.4}$$

for all Boolean functions $f_i$ on $X_i$ $(i = 1, 2, \dots, n)$. Lemma 3.3 now implies that

$$|||\Psi_k|||^* \leqslant k!\binom{n+k}{k} C_1 C_2^k \mathop{\mathbf{E}}_{|S|=n-k}\left[\prod_{i\in S}|||\psi_i|||^*\right],$$

which settles (4.2) in view of (4.3). ∎

We now prove a direct product theorem, again in the context of an arbitrary norm. More specifically, the theorem places a lower bound on the norm of any $(\sigma, m)$-approximant for a given set of functions, as formalized in the following definition.

DEFINITION 4.2. Fix finite sets $X_1, X_2, \dots, X_n$ and a norm $|||\cdot|||$ on $\mathbb{R}^{\prod X_i}$. For a (possibly partial) Boolean function $g_i$ on $X_i$ $(i = 1, 2, \dots, n)$, let

$$|||g_1, g_2, \dots, g_n, \sigma, m||| = \min_{\{\phi_z\}} \max_{z\in\{-1,+1\}^n} |||\phi_z|||,$$

where the minimum is over all $(\sigma, m)$-approximants $\{\phi_z\}$ of $(g_1, g_2, \dots, g_n)$.

THEOREM 4.3. *Fix finite sets $X_1, X_2, \dots, X_n$. Let $\mathbb{R}^{X_1}, \mathbb{R}^{X_2}, \dots, \mathbb{R}^{X_n}$, and $\mathbb{R}^{\prod X_i}$ be normed by $|||\cdot|||$, with $C_1, C_2$ defined by* (3.10), (3.11). *Fix a* (*possibly partial*) *Boolean function $g_i$ on $X_i$ $(i = 1, 2, \dots, n)$. Then for all $\epsilon, \sigma \in (0, 1)$, and all nonnegative integers $k, \ell, m$ with $k + \ell \leqslant n$, one has:*

$$|||g_1, g_2, \dots, g_n, \sigma, m||| \geqslant$$
$$\frac{\prod_{i=1}^{n} |||g_i|||_{1-\epsilon}}{\mathop{\mathbf{E}}_{|S|=k+\ell}\left[\prod_{i\in S}|||g_i|||_{1-\epsilon}\right]} \cdot \frac{\sigma - 2^{-\frac{\alpha\ell^2}{n}+m+1} - 2\binom{n}{k+1}\frac{(\frac{1}{2}\epsilon)^{k+1}}{(1-\frac{1}{2}\epsilon)^n}}{\left(\frac{2^n\,\epsilon^{n-k-\ell}}{(1-\frac{1}{2}\epsilon)^n}\binom{n+k}{k}\binom{n}{\leqslant\ell}\right)^{1/2} C_1 C_2^{k+\ell}},$$

*where $\alpha > 0$ is the absolute constant from Theorem* 3.5.

*Proof.* By Corollary 2.2, for all $i$ there exists $\psi_i\colon X_i \to \mathbb{R}$ that obeys (4.1). Clearly the expression in braces in (4.1) is positive for all $i$. By homogeneity, we may assume that $\|\psi_i\|_1 = 1$. Fix a $(\sigma, m)$-approximant for $(g_1, g_2, \dots, g_n)$ and define $\Psi_k, \Phi_\ell\colon \prod X_i \to \mathbb{R}$ as in Lemma 3.7. Define $\Psi_{k,\ell,z}\colon \prod X_i \to \mathbb{R}$ by

$$\Psi_{k,\ell,z}(\dots, x_i, \dots) = \Psi_k(\dots, x_i, \dots) q_\ell(\dots, z_i f_i(x_i), \dots) \prod_{i=1}^{n} z_i,$$

where $f_1, f_2, \dots, f_n$ are as defined in Lemma 3.7 and $q_\ell$ is the degree-$\ell$ polynomial guaranteed in Corollary 3.6.

CLAIM 4.4. *For each $z \in \{-1,+1\}^n$,*

$$|||\Psi_{k,\ell,z}|||^* \leqslant k!\binom{n+k}{k}\binom{n}{\leqslant \ell}^{1/2} C_1 C_2^{k+\ell} \mathop{\mathbf{E}}_{|S|=n-k-\ell}\left[\prod_{i\in S}|||\psi_i|||^*\right].$$

*Proof.* Let $\mathscr{C}_0, \mathscr{C}_1, \dots, \mathscr{C}_n$ be as defined in Lemma 3.3. Let $p_k$ be the degree-$k$ polynomial defined in Lemma 3.1. We have

$$p_k(\dots, f_i(x_i)\operatorname{sgn}\psi_i(x_i), \dots) \in k!\binom{n+k}{k}\mathscr{C}_k,$$

$$q_\ell(\dots, z_i f_i(x_i), \dots) \in \binom{n}{\leqslant \ell}^{1/2}\mathscr{C}_\ell.$$

The first of these memberships was shown earlier in (4.4), and the second is immediate from (3.19) and the symmetry of $q_\ell$. The product of these two functions therefore lies in $k!\binom{n+k}{k}\binom{n}{\leqslant\ell}^{1/2}\mathscr{C}_{k+\ell}$. Since

$$\Psi_{k,\ell,z}(\dots, x_i, \dots) = p_k(\dots, f_i(x_i)\operatorname{sgn}\psi_i(x_i), \dots) q_\ell(\dots, z_i f_i(x_i), \dots)\prod_{i=1}^n z_i\psi_i(x_i),$$

by definition, the proof complete by Lemma 3.3. ∎

Note that (4.1) forces

$$|||\psi_i|||^* \leqslant \frac{\epsilon\|\psi_i\|_1}{|||g_i|||_{1-\epsilon}} = \frac{\epsilon}{|||g_i|||_{1-\epsilon}}, \qquad i = 1, 2, \dots, n. \tag{4.5}$$

Now,

$$\begin{aligned}\langle \Phi_\ell, \Psi_k\rangle &= \sum_{z\in\{-1,+1\}^n}\langle \phi_z, \Psi_{k,\ell,z}\rangle \\ &\leqslant \sum_{z\in\{-1,+1\}^n} |||\phi_z|||\,|||\Psi_{k,\ell,z}|||^* \\ &\leqslant k!\,2^n\binom{n+k}{k}\binom{n}{\leqslant\ell}^{1/2} C_1C_2^{k+\ell}\mathop{\mathbf{E}}_{|S|=n-k-\ell}\left[\prod_{i\in S}|||\psi_i|||^*\right]\max_z|||\phi_z|||,\end{aligned}$$

where the last step follows by Claim 4.4. Now (4.5) shows that $\langle\Phi_\ell, \Psi_k\rangle$ cannot exceed

$$k!\,2^n\epsilon^{n-k-\ell}\binom{n+k}{k}\binom{n}{\leqslant\ell}^{1/2} C_1C_2^{k+\ell}\mathop{\mathbf{E}}_{|S|=n-k-\ell}\left[\prod_{i\in S}\frac{1}{|||g_i|||_{1-\epsilon}}\right]\max_z|||\phi_z|||.$$

In view of (3.20) and the fact that the theorem is void for $\sigma < 2^{-\alpha\ell^2/n+m+1}$, we obtain the claimed lower bound on $\max|||\phi_z|||$. ∎

**4.2. XOR lemmas.** We now specialize the above results to the $\gamma_2$ norm and quantum communication complexity. For this, we recall a multiplicative property of the dual norm $\gamma_2^*$, established by Cleve, Slofstra, Unger, and Upadhyay [17].

THEOREM 4.5 (Cleve et al.). *For all real matrices* $A, B$,

$$\gamma_2^*(A \otimes B) = \gamma_2^*(A)\gamma_2^*(B).$$

Theorem 4.5 was revisited more recently by Lee, Shraibman, and Špalek [37, Thm. 17], who additionally showed multiplicativity for the primal norm $\gamma_2$. For our purposes, only the upper bound part of Theorem 4.5 is needed. We have:

THEOREM 4.6. *For all* (*possibly partial*) *sign matrices* $F_1, F_2, \ldots, F_n$ *and all sufficiently small constants* $\epsilon > 0$,

$$\gamma_{2,1-\epsilon^{n/101}}\left(\bigotimes_{i=1}^{n} F_i\right) \geqslant \min_{|S|=\lceil 0.99n\rceil}\left\{\prod_{i\in S}\gamma_{2,1-\epsilon}(F_i)\right\}.$$

*Proof.* By Theorem 4.5 and Fact 2.3(iv), the norm $|||\cdot||| = \gamma_2$ satisfies (3.10) with $C_1 \leqslant 1$ and (3.11) with $C_2 \leqslant 1$. Hence, the result follows from Theorem 4.1 by letting $k = \lfloor 0.01n\rfloor$ and $\delta = 1-\epsilon^{n/101}$. ∎

Theorem 4.6 gives the desired XOR lemma for quantum communication. We will now show how to improve the dependence on the constant $\epsilon$ in the more interesting case of *total* communication problems. A base case in our analysis is given by:

PROPOSITION 4.7. *Let* $F_1, F_2, \ldots, F_n$ *be sign matrices, each of rank at least* 2. *Then*

$$\gamma_{2,1-\epsilon}\left(\bigotimes_{i=1}^{n} F_i\right) \geqslant \epsilon 2^{n/2} \qquad (0 \leqslant \epsilon \leqslant 1). \tag{4.6}$$

*Proof.* Let $H$ be given by (2.6). We claim that each $F_i$ contains some signature scaling of $H$ as a submatrix. To see this, signature scale $F_i$ such that the first row and first column feature only $+1$ entries and conclude from $\operatorname{rk} F_i > 1$ the existence of a $-1$ entry elsewhere in the resulting matrix. Now (4.6) is immediate from Fact 2.3 (i), (ii), (xi). ∎

We have:

THEOREM 4.8. *For every sign matrix* $F$,

$$\gamma_{2,1-(\frac{3}{4})^n}(F^{\otimes n}) \geqslant \left(\gamma_{2,\frac{1}{4}}(F)\right)^{\Omega(n)}. \tag{4.7}$$

*Proof.* In the trivial case when $F$ has rank 1, Fact 2.3 (i), (x) gives $\gamma_{2,1-(3/4)^n}(F^{\otimes n}) = (3/4)^n$ and $\gamma_{2,1/4}(F) = 3/4$, proving (4.7).

In the remainder of the proof, we assume that $\operatorname{rk} F > 1$. By Theorem 4.5 and Fact 2.3(iv), the norm $|||\cdot||| = \gamma_2$ satisfies (3.10) with $C_1 \leqslant 1$ and (3.11) with $C_2 \leqslant 1$. As a result, letting $k = \lfloor 0.96n\rfloor$, $\epsilon = 3/4$, $\delta = 1-(3/4)^n$, and $f_1 = f_2 = \cdots = f_n = F$

in Theorem 4.1 yields

$$\gamma_{2,1-(\frac{3}{4})^n}(F^{\otimes n}) > \gamma_{2,\frac{1}{4}}(F)^{n/25} \cdot 19^{-n}.$$

Proposition 4.7 with $\operatorname{rk} F > 1$ gives

$$\gamma_{2,1-(\frac{3}{4})^n}(F^{\otimes n}) \geqslant \left(\frac{3}{2\sqrt{2}}\right)^n.$$

One now obtains (4.7) as a geometric mean of these two lower bounds: for small $\beta > 0$,

$$\gamma_{2,1-(\frac{3}{4})^n}(F^{\otimes n}) \geqslant \left\{\gamma_{2,\frac{1}{4}}(F)^{n/25} \cdot 19^{-n}\right\}^\beta \left\{\left(\frac{3}{2\sqrt{2}}\right)^n\right\}^{1-\beta} \geqslant \gamma_{2,\frac{1}{4}}(F)^{\beta n/25}. \qquad \square$$

Theorem 4.8 readily generalizes to $n$ distinct sign matrices:

THEOREM 4.9. *Fix sign matrices $F_1, F_2, \ldots, F_n$, each of rank at least 2. Then*

$$\gamma_{2,1-(\frac{3}{4})^n}\left(\bigotimes_{i=1}^n F_i\right) \geqslant \left(\min_{|S|=\lceil n/25\rceil}\left\{\prod_{i\in S}\gamma_{2,1/4}(F_i)\right\}\right)^{\Theta(1)}. \tag{4.8}$$

*In particular,*

$$\gamma_{2,1-2^{-\Omega(n)}}\left(\bigotimes_{i=1}^n F_i\right) \geqslant \left(\min_{|S|=\lceil 0.99n\rceil}\left\{\prod_{i\in S}\gamma_{2,1/4}(F_i)\right\}\right)^{\Theta(1)}. \tag{4.9}$$

*Proof.* By Theorem 4.5 and Fact 2.3(iv), the norm $|||\cdot||| = \gamma_2$ satisfies (3.10) with $C_1 \leqslant 1$ and (3.11) with $C_2 \leqslant 1$. As a result, letting $k = \lfloor 0.96n\rfloor$, $\epsilon = 3/4$, and $\delta = 1-(3/4)^n$ in Theorem 4.1 yields

$$\gamma_{2,1-(\frac{3}{4})^n}\left(\bigotimes_{i=1}^n F_i\right) > 19^{-n}\min_{|S|=\lceil n/25\rceil}\left\{\prod_{i\in S}\gamma_{2,1/4}(F_i)\right\}.$$

Proposition 4.7 gives an alternate lower bound of $(3\sqrt{2}/4)^n$. One now obtains (4.8) as a geometric mean of these two lower bounds, as was done in the proof of Theorem 4.8. Finally, (4.8) trivially implies (4.9). $\square$

This establishes the XOR results in Theorems 1.1 and 1.3 of the Introduction.

**4.3. Direct product theorems.** We will now derive direct product theorems for quantum communication, corresponding to the XOR lemmas just obtained. Recall that the symbol $Q^*_{1-\sigma,m}(F_1, F_2, \ldots, F_n)$ stands for the least cost of a quantum protocol that solves with probability $\sigma$ at least $n-m$ of the communication problems $F_1, F_2, \ldots, F_n$. The meaningful case is when the ratio $m/n$ is a sufficiently small constant. In this setting, a protocol that simply outputs a random answer $(z_1, z_2, \ldots, z_n) \in \{-1,+1\}^n$ without any communication achieves error probability $1 - 2^{-n}\binom{n}{\leqslant m} = 1 - 2^{-\Omega(n)}$. All communication lower bounds below allow the protocol to err with probability $1 - 2^{-\Omega(n)}$.

For functions $F_i\colon X_i \times Y_i \to \{-1,+1\}$, $i = 1, 2, \ldots, k$, and a gadget $g\colon \{-1,+1\}^k \to \{-1,+1\}$, the symbol $g(F_1, \ldots, F_k)$ stands as usual for their composition, which is a function $\prod X_i \times \prod Y_i \to \{-1,+1\}$. We will mostly be interested in $g = \wedge$ and $g = \oplus$ below. We will usually view $g(F_1, \ldots, F_k)$ as a sign matrix, with rows indexed by elements of $\prod X_i$ and columns indexed by elements of $\prod Y_i$. For (possibly partial) sign matrices $F_1, \dots, F_n$, define $\gamma_2(F_1, \dots, F_n, \sigma, m)$ to be $|||F_1, \dots, F_n, \sigma, m|||$ with $|||\cdot|||$ taken to be the $\gamma_2$ norm on the matrix family $\mathbb{R}^{\prod X_i \times \prod Y_i}$.

PROPOSITION 4.10. *For all* (*possibly partial*) *sign matrices* $F_1, \dots, F_n$,

$$2^{Q^*_{1-\sigma,m}(F_1,\dots,F_n)} \geqslant \gamma_2(F_1, \dots, F_n, \sigma, m).$$

*Proof.* For a protocol $\Pi$ with cost $c$ that solves with probability $\sigma$ at least $n-m$ of the problems $F_1, \dots, F_n$, define $\phi_z(x_1, \dots, x_n, y_1, \dots, y_n) = \mathbf{P}[\Pi(x_1, \dots, x_n, y_1, \dots, y_n) = z]$, where the probability is taken over the operation of the protocol on a fixed input. Then $\{\phi_z\}$ is a $(\sigma, m)$-approximant for $(F_1, \dots, F_n)$. Viewed as an element of $\mathbb{R}^{\prod X_i \times \prod Y_i}$, each $\phi_z$ is the matrix of acceptance probabilities of a quantum protocol with one-bit output and cost $c$ (namely, the quantum protocol that accepts if and only if $\Pi$ outputs $z$). Thus, $\gamma_2(\phi_z) \leqslant 2^c$ by Theorem 2.6. ∎

Recall from Theorem 4.5 and Fact 2.3(iv) that the norm $|||\cdot||| = \gamma_2$ satisfies (3.10) with $C_1 \leqslant 1$ and (3.11) with $C_2 \leqslant 1$. We will use this fact without further mention whenever we invoke our main technical tool here, Theorem 4.3. We have:

THEOREM 4.11. *Fix* (*possibly partial*) *sign matrices* $F_1, F_2, \dots, F_n$. *Then for a sufficiently small constant* $\epsilon > 0$,

$$Q^*_{1-2^{-\epsilon n},\epsilon n}(F_1, F_2, \dots, F_n) \geqslant \epsilon \min_{|S|=\lceil 0.99n \rceil} \left\{ \sum_{i \in S} \log\left(\gamma_{2,1-\epsilon}(F_i)\right) \right\}.$$

*Proof.* A protocol that solves $(F_1, F_2, \dots, F_n)$ with probability 0.99 can solve each $F_i$ individually with probability 0.99. Hence, for $n$ up to any given constant, the theorem follows trivially from Theorem 2.7 by choosing $\epsilon > 0$ correspondingly small. For $n$ larger than a certain constant, the theorem follows by taking $k = \ell = \lfloor 0.005n \rfloor$, $m = \lfloor \epsilon n \rfloor$, and $\sigma = 2^{-\epsilon n}$ in Theorem 4.3 and applying Proposition 4.10. ∎

Theorem 4.11 gives the desired direct product theorem for quantum communication. As we did for XOR lemmas, we will now take a closer look at the more interesting case of total functions, improving several constants. The base case in our analysis is given by the following statement.

LEMMA 4.12. *Let* $\beta > 0$ *be a sufficiently small absolute constant. Then:*

(i) $Q^*_{1-2^{-\beta n},\beta n}(F_1, \dots, F_n) > \beta n$ *for any order-*2 *Hadamard matrices* $F_1, \dots, F_n$.

(ii) $Q^*_{1-2^{-\beta n},\beta n}(F_1, \dots, F_n) > \beta n$ *for any sign matrices* $F_1, \dots, F_n$ *of rank at least* 2.

*Proof.* (i) Let $K \geqslant 1$ be a sufficiently large integer constant and $H_1, \dots, H_r$ Hadamard matrices of order $2^K$. Then $\gamma_{2,1-\epsilon}(H_i) = \epsilon 2^{K/2}$ for each $i$, by Fact 2.3(xi). As a result, choosing a sufficiently small absolute constant $\beta' > 0$ and letting $n = r$, $k = \lfloor 0.96r \rfloor$, $\ell = \lfloor 0.01r \rfloor$, $\epsilon = 3/4$, and $\sigma = (3/4)^r + 2^{-\alpha \ell^2/r + \beta' r + 1}$ in Theorem 4.3 give $\gamma_2(H_1, \dots, H_r, \beta' r, 2^{-\beta' r}) > 2^{\beta' r}$, whence $Q^*_{1-2^{-\beta' r},\beta' r}(H_1, \dots, H_r) > \beta' r$ by Proposition 4.10.

Now, given order-2 Hadamard matrices $F_1, \dots, F_n$, let $r = \lfloor n/K \rfloor$ and consider the order-$2^K$ Hadamard matrices $H_i = F_{K(i-1)+1} \otimes F_{K(i-1)+2} \otimes \cdots \otimes F_{Ki}$ for $i = 1, 2, \dots, r$. We have $Q^*_{1-2^{-\beta' r}, \beta' r}(H_1, \dots, H_r) > \beta' r$ by above, which proves the claim since trivially $Q^*_{1-\sigma,\beta' r}(F_1, \dots, F_n) \geqslant Q^*_{1-\sigma,\beta' r}(H_1, \dots, H_r)$ for all $\sigma$.

(ii) As argued earlier in Proposition 4.7, each $F_i$ contains a Hadamard matrix of order 2. The proof is now complete by (i). ∎

Using the previous lemma, we will now derive the sought direct product theorems.

THEOREM 4.13. *Fix sign matrices $F_1, F_2, \dots, F_n$, each of rank at least 2. Then for some absolute constant $\beta > 0$,*

$$Q^*_{1-2^{-\beta n},\beta n}(F_1, F_2, \dots, F_n) \geqslant \beta \min_{|S|=\lceil 0.99n \rceil} \left\{ \sum_{i \in S} \log\left(\gamma_{2,\frac{1}{4}}(F_i)\right) \right\}. \tag{4.10}$$

*Proof.* As explained in the proof of Theorem 4.11, we may assume that $n$ is larger than a constant of our choice. Choosing a sufficiently small absolute constant $\beta' > 0$, letting $k = \lfloor 0.96n \rfloor$, $\ell = \lfloor 0.01n \rfloor$, $\epsilon = 3/4$, and $\sigma = (3/4)^n + 2^{-\alpha\ell^2/n + \beta' n + 1}$ in Theorem 4.3, and invoking Proposition 4.10 give

$$Q^*_{1-2^{-\beta' n},\beta' n}(F_1, \dots, F_n) \geqslant -n \log 18 + \min_{|S|=\lceil 0.03n \rceil} \left\{ \sum_{i \in S} \log\left(\gamma_{2,1/4}(F_i)\right) \right\},$$

for all $n$ larger than a certain constant. Lemma 4.12(ii) gives an alternate lower bound of

$$Q^*_{1-2^{-\beta'' n},\beta'' n}(F_1, \dots, F_n) > \beta'' n$$

for some absolute constant $\beta'' > 0$. Taking a weighted arithmetic average of these bounds yields

$$Q^*_{1-2^{-\beta n},\beta n}(F_1, \dots, F_n) \geqslant \beta \min_{|S|=\lceil 0.03n \rceil} \left\{ \sum_{i \in S} \log\left(\gamma_{2,1/4}(F_i)\right) \right\}$$

for a constant $\beta > 0$, which is logically equivalent to (4.10). ∎

The lower bound (4.10) establishes the direct product results in Theorems 1.1 and 1.3 of the Introduction.

The XOR of any sign matrices $F_1, F_2, \dots, F_n$ of rank 1 is computable with a single bit of communication. In particular, it is meaningless to speak of an XOR lemma in that case. Direct product theorems, however, remain meaningful even for rank-1 matrices. While this case is of minor interest, we include its simple solution for completeness. Call a matrix *column-constant* if its columns are all identical, and similarly for *row-constant*.

THEOREM 4.14. *Fix rank-1 sign matrices $F_1, F_2, \dots, F_n$, of which $c$ are not column-constant and $r$ are not row-constant. Then for some absolute constant $\beta > 0$,*

$$Q^*_{1-2^{-\beta t},\beta t}(F_1, F_2, \dots, F_n) \geqslant \beta t,$$

*where $t = \min\{r, c\}$. The bound is tight in that a one-way classical deterministic protocol can simultaneously solve $F_1, F_2, \dots, F_n$ using $t$ bits of communication.*

*Proof.* Since each matrix has rank 1, there can be at most two distinct rows and at most two distinct columns per matrix. This gives the deterministic upper bound: if $r \leqslant c$, the row player identifies his entire input by sending one bit for each of the matrices that are not row-constant, and similarly if $c < r$.

The quantum lower bound follows from Lemma 4.12(ii). Namely, select disjoint subsets $I, J \subset \{1, 2, \ldots, n\}$ with $|I| = |J| \geqslant \min\{\lfloor r/2 \rfloor, \lfloor c/2 \rfloor\}$ such that none of $\{F_i : i \in I\}$ are column-constant and none of $\{F_j : j \in J\}$ are row-constant. Then for all $i \in I$ and $j \in J$, the sign matrix $F_i \wedge F_j$ has rank at least 2. As a result, any protocol for $(F_1, F_2, \ldots, F_n)$ can be turned into a protocol for $|I| = |J|$ sign matrices of rank at least 2, with the same performance guarantees. ∎

Together, Theorems 4.13 and 4.14 give direct product theorems for arbitrary $n$-tuples of sign matrices: one partitions the given sign matrices $F_1, F_2, \ldots, F_n$ into a group of rank 1 and a group of rank greater than 1, invokes the corresponding theorem for each group, and takes the maximum of the lower bounds thus obtained.

As a final remark, Theorems 4.13 and 4.14 can be strengthened with respect to the protocol's error probability by providing a sharper approximant than what is guaranteed in Theorem 3.5. We will illustrate this point by deriving, for an arbitrarily small constant $\xi > 0$, a strong direct product theorem for protocols with error probability $1-2^{-(1-\xi)n}$. This bound essentially matches the error probability $1-2^{-n}$ achieved by a trivial, communication-free protocol. The approximant in Theorem 3.5 is no longer sufficient for this purpose, and we obtain the necessary statement via a more careful analysis of an earlier approximant due to Kahn, Kalai, and Linial [27].

LEMMA 4.15. *Let $\xi \in (0, 1)$ be a given constant. Then Theorem 3.5 holds with $m = 0$ and $\alpha = 1 - \xi$ provided that $\ell \geqslant (1 - \xi')n$, where $\xi' = \xi'(\xi) > 0$ is another constant.*

*Proof.* Following [27], we put $r = \lfloor \ell/2 \rfloor$ and define

$$Q_\ell(t) = \frac{1}{r!\,r!}\binom{n}{r}^{-1} \prod_{i=0}^{r-1}(t - i - 1)(t - n + i).$$

Then

$$\begin{aligned} Q_\ell(t) &= 0, && t \in \{1, \ldots, r\} \cup \{n - r + 1, \ldots, n\}, \\ |Q_\ell(t)| &\leqslant \binom{n-r}{r}^2 \binom{n}{r}^{-1}, && t \in \{r + 1, \ldots, n - r\}, \\ Q_\ell(0) &= 1. \end{aligned}$$

Thus,

$$\max_{t=1,\ldots,n} |Q_\ell(t)| \leqslant 2^{-n\left\{H\left(\frac{r}{n}\right) - 2H\left(\frac{r}{n-r}\right)\right\}(1-o(1))}, \tag{4.11}$$

where $H$ is the binary entropy function. In particular, for any $\xi > 0$, the right member of (4.11) is bounded by $2^{-(1-\xi)n}$ provided that one has $\ell/n \geqslant 1 - \xi'$ for some constant $\xi' = \xi'(\xi) > 0$. ∎

We now prove the desired direct product theorem.

THEOREM 4.16. *Let $\xi > 0$ be a constant and $F_1, F_2, \dots, F_n$ sign matrices. Then*

$$Q^*_{1-2^{-(1-\xi)n}}(F_1, \dots, F_n) \geqslant \min_{|S|=\lfloor \zeta n \rfloor} \left\{ \sum_{i \in S} \log\left(\gamma_{2,1-\epsilon}(F_i)\right) \right\}$$

*for a constant $\zeta = \zeta(\xi) > 0$ and all sufficiently small constants $\epsilon > 0$.*

*Proof.* By Lemma 4.15, for $m = 0$, the constant $\alpha \in (0, 1)$ in Theorems 3.5 and 4.3 can be taken to be arbitrarily close to 1 provided that the ratio $\ell/n$ is larger than a corresponding constant in (0, 1). Thus, the claim follows by taking $|||\cdot||| = \gamma_2$, $m = 0$, $k = \lfloor (n-\ell)/2 \rfloor$, $\sigma = 2^{-(1-\xi)n}$, and $\ell = \lfloor (1-\xi')n \rfloor$ for a small enough constant $\xi' = \xi'(\xi) > 0$ in Theorem 4.3 and applying Proposition 4.10. ∎

**4.4. Direct sum theorems.** In a final result on quantum communication complexity, we prove a direct sum property for approximation in the $\gamma_2$ norm. In view of Theorem 2.7, this translates to a direct sum theorem for quantum communication whenever the original lower bounds were obtained by the generalized discrepancy method. As explained in the Introduction, this result is incomparable with the direct product theorems derived in the previous subsection. We start with a technical fact.

PROPOSITION 4.17. *Fix nonnegative reals $a_1, a_2, \dots, a_n$, with $a = \max\{a_1, a_2, \dots, a_n\}$. Partition $\{1, 2, \dots, n\} = S_1 \cup S_2 \cup S_3 \cup \dots$, where $S_i = \{j : a_j \in (2^{-i}a, 2^{-i+1}a]\}$. Then*

$$\sum_{i:|S_i| \geqslant i/8} |S_i| \min\{a_j : j \in S_i\} \geqslant \frac{1}{4} \sum_{i=1}^{n} a_i.$$

*Proof.* Define shorthands $M_i = \max\{a_j : j \in S_i\}$ and $m_i = \min\{a_j : j \in S_i\}$. We have

$$\begin{aligned} \sum_{i:|S_i| \geqslant i/8} |S_i|\, m_i &\geqslant \sum_{i \geqslant 1} \left(|S_i| - \frac{i}{8}\right) m_i \geqslant \sum_{i \geqslant 1} \left(|S_i| - \frac{i}{8}\right) \frac{a}{2^i} \\ &\geqslant a \left( \sum_{i \geqslant 1} \frac{|S_i|}{2^i} - \frac{1}{4} \right) \geqslant \frac{a}{2} \sum_{i \geqslant 1} \frac{|S_i|}{2^i}, \end{aligned}$$

where the last step follows because $|S_1| \geqslant 1$. The final expression is bounded from below by $\frac{1}{4} \sum_{i \geqslant 1} |S_i|\, M_i \geqslant \frac{1}{4} \sum_{i=1}^{n} a_i$. ∎

We have:

THEOREM 4.18. *For a sufficiently small constant $\epsilon > 0$ and arbitrary* (*possibly partial*) *sign matrices $F_1, F_2, \dots, F_n$,*

$$\gamma_{2,1/4}\left( \bigotimes_{i=1}^{n} F_i \right) \geqslant \left( \prod_{i=1}^{n} \gamma_{2,1-\epsilon}(F_i) \right)^{\Theta(1)}. \tag{4.12}$$

*Proof.* Let $\epsilon > 0$ be sufficiently small in the sense of Theorem 4.6. We may assume that $\gamma_{2,1-\epsilon}(F_i) \geqslant 1$ for all $i$ since any offending matrices decrease the right member of (4.12) without decreasing the left. Let $a_i = \ln\{\gamma_{2,1-\epsilon}(F_i)\}$ and $a = \max\{a_1, a_2, \ldots, a_n\}$. Define $S_1, S_2, S_3, \ldots$ as in Proposition 4.17. Theorem 4.6 shows that

$$\gamma_{2,1-M^{-k}}\left(\bigotimes_{i=1}^{k} G_i\right) \geqslant \left(\min_{i=1,2,\ldots,k}\{\gamma_{2,1-\epsilon}(G_i)\}\right)^{\lceil 0.99k\rceil} \tag{4.13}$$

for all $k$ and all (possibly partial) sign matrices $G_1, G_2, \ldots, G_k$, where $M = M(\epsilon) > 1$ is a constant that can be made as large as desired by choosing $\epsilon > 0$ sufficiently small. For large enough $M$, we have

$$\delta = 2\prod_{i=1}^{\infty}\left\{1 - \frac{1}{M^{i/8}}\right\} - 1 > \frac{1}{4}$$

and

$$\begin{aligned}
\gamma_{2,\delta}\left(\bigotimes_{i=1}^{n} F_i\right) &\geqslant \gamma_{2,\delta}\left(\bigotimes_{i:|S_i|\geqslant i/8}\ \bigotimes_{j\in S_i} F_j\right) && \text{by Fact 2.3(ii)}\\
&\geqslant \prod_{i:|S_i|\geqslant i/8} \gamma_{2,1-M^{-|S_i|}}\left(\bigotimes_{j\in S_i} F_j\right) && \text{by Lemma 3.8 with } C_1 = 1\\
&\geqslant \exp\left(0.99\sum_{i:|S_i|\geqslant i/8} |S_i|\min\{a_j : j\in S_i\}\right) && \text{by (4.13)}\\
&\geqslant \left(\prod_{i=1}^{n}\gamma_{2,1-\epsilon}(F_i)\right)^{1/5} && \text{by Proposition 4.17.}
\end{aligned}$$

The above appeal to Lemma 3.8 with $C_1 = 1$ is legitimate by Theorem 4.5. ∎

In the remainder of this subsection, we will take a more careful look at the case of total functions and improve the dependence on $\epsilon$. For this, we need a standard error-reduction property for uniform approximation out of the unit ball of $\gamma_2$:

FACT 4.19. *Fix a sign matrix $F$ with* $\operatorname{rk} F \geqslant 2$ *and a constant* $\epsilon \in (0, 1/4)$. *Then*

$$\gamma_{2,\frac{1}{4}}(F) \geqslant \gamma_{2,\epsilon}(F)^{\epsilon'}$$

*for some constant* $\epsilon' = \epsilon'(\epsilon) > 0$.

This fact follows easily by applying an approximating polynomial to the matrix entries, as was done in earlier papers, e.g., [3], [33]. Details follow.

*Proof of Fact* 4.19. Take a real matrix $A$ with $\|F - A\|_\infty \leqslant 1/4$ and $\gamma_2(A) = \gamma_{2,1/4}(F)$. By basic approximation theory [44], there is a univariate polynomial $p(t) = \sum_{i=1}^{d} a_i t^i$ of degree $d = O(1)$ that sends $[-5/4, -3/4] \to [-1-\epsilon, -1+\epsilon]$ and $[3/4, 5/4] \to [1-\epsilon, 1+\epsilon]$. Then $\|F - B\|_\infty \leqslant \epsilon$ for

$$B = \sum_{i=1}^{d} a_i \underbrace{A \circ A \circ \cdots \circ A}_{i \text{ times}}.$$

Now Fact 2.3(xiii) gives $\gamma_2(B) \leqslant \sum_{i=1}^{d} |a_i|\, \gamma_2(A)^i$, which is bounded by $\gamma_2(A)^{\Theta(1)}$ since $\gamma_2(A) = \gamma_{2,1/4}(F) > 1.06$ by Proposition 4.7. ∎

We now arrive at the desired direct sum theorem.

THEOREM 4.20. *For any sign matrices $F_1, F_2, \ldots, F_n$,*

$$\gamma_{2,1/4}\left(\bigotimes_{i=1}^{n} F_i\right) \geqslant \left(\prod_{i=1}^{n} \gamma_{2,1/4}(F_i)\right)^{\Theta(1)}. \tag{4.14}$$

*Proof.* We may discard any matrices among $F_1, F_2, \ldots, F_n$ that have rank 1: by Fact 2.3 (i), (iii), their presence does not affect the left member of (4.14) but by Fact 2.3(x) decreases the right member. From now on, we will assume that $F_1, F_2, \ldots, F_n$ have rank at least 2.

Let $a_i = \ln\{\gamma_{2,1/4}(F_i)\}$ and $a = \max\{a_1, a_2, \ldots, a_n\}$. Then $a_i > 0$ for all $i$, by Proposition 4.7. Define $S_1, S_2, S_3, \ldots$ as in Proposition 4.17. Letting

$$\delta = \prod_{i=1}^{\infty} \left\{1 - \left(\frac{3}{4}\right)^{i/8}\right\} = \Omega(1),$$

we have:

$$\begin{aligned}
\gamma_{2,\delta}\left(\bigotimes_{i=1}^{n} F_i\right) &\geqslant \gamma_{2,\delta}\left(\bigotimes_{i:|S_i|\geqslant i/8}\ \bigotimes_{j\in S_i} F_j\right) && \text{by Fact 2.3(ii)}\\
&\geqslant \prod_{i:|S_i|\geqslant i/8} \gamma_{2,1-\left(\frac{3}{4}\right)^{|S_i|}}\left(\bigotimes_{j\in S_i} F_j\right) && \text{by Lemma 3.8 with } C_1 = 1\\
&\geqslant \exp\left\{\Omega\left(\sum_{i:|S_i|\geqslant i/8} |S_i| \min\{a_j : j \in S_i\}\right)\right\} && \text{by Theorem 4.9}\\
&\geqslant \left(\prod_{i=1}^{n} \gamma_{2,1/4}(F_i)\right)^{\Theta(1)} && \text{by Proposition 4.17.}
\end{aligned}$$

The above appeal to Lemma 3.8 with $C_1 = 1$ is legitimate by Theorem 4.5. By Fact 4.19, the proof is complete. ∎

This establishes Theorem 1.2 from the Introduction.

## 5. QUANTUM QUERY COMPLEXITY

This section is devoted to our results on query complexity. We prove XOR lemmas, direct product theorems, and direct sum theorems for polynomial approximation and thereby obtain the claimed consequences for quantum query complexity.

**5.1. XOR lemmas.** We start with an XOR lemma for polynomial approximation. The development here closely parallels our earlier proof of an XOR lemma for norm-based computation.

THEOREM 5.1. *Fix a (possibly partial) Boolean function $g_i$ on $X_i$ $(i = 1, 2, \ldots, n)$, for finite sets $X_1, X_2, \ldots, X_n \subset \mathbb{R}^m$. Then for every $\epsilon \in (0, 1)$ and all $k = 0, 1, \ldots, n-1$,*

$$\deg_{1-2\binom{n}{k+1}\frac{(\epsilon/2)^{k+1}}{(1-\epsilon/2)^n}}\left(\bigotimes_{i=1}^n g_i\right) \geqslant \min_{|S|=n-k}\left\{\sum_{i\in S}\deg_{1-\epsilon}(g_i)\right\}.$$

*Proof.* For $i = 1, 2, \ldots, n$, Theorem 2.4 provides $\psi_i \colon X_i \to \mathbb{R}$ that obeys (3.5), (3.6), and

$$\sum_{x_i\in X_i}\psi_i(x_i)p(x_i) = 0 \tag{5.1}$$

for every polynomial $p$ of degree less than $\deg_{1-\epsilon}(g_i)$. Define $\Psi_k \colon \prod X_i \to \mathbb{R}$ as in Lemma 3.2. Then (5.1) shows that $\sum_{\prod X_i}\Psi_k(\ldots, x_i, \ldots)p(\ldots, x_i, \ldots) = 0$ for every polynomial $p$ of degree less than $\min_{|S|=n-k}\left\{\sum_{i\in S}\deg_{1-\epsilon}(g_i)\right\}$. By (3.7),

$$\sum_{x\in\prod \operatorname{dom} g_i}\Psi_k(x)\prod_{i=1}^n g_i(x_i) - \sum_{x\notin\prod \operatorname{dom} g_i}|\Psi_k(x)| - \left\{1 - 2\binom{n}{k+1}\frac{\left(\frac{1}{2}\epsilon\right)^{k+1}}{\left(1-\frac{1}{2}\epsilon\right)^n}\right\}\|\Psi_k\|_1 > 0.$$

By Theorem 2.4, the proof is complete. ∎

COROLLARY 5.2. *Fix a (possibly partial) Boolean function $g_i$ on $X_i$ $(i = 1, 2, \ldots, n)$, for some finite sets $X_1, X_2, \ldots, X_n \subset \mathbb{R}^m$. Then for every constant $\beta > 0$,*

$$\deg_{1-\beta^n}\left(\bigotimes_{i=1}^n g_i\right) \geqslant \Omega\left(\min_{|S|=\lceil 0.99n\rceil}\left\{\sum_{i\in S}\deg_{1/3}(g_i)\right\}\right).$$

*Proof.* Recall from (2.9) that $\deg_\epsilon(g) = \Theta(\deg_{\epsilon'}(g))$ for any partial or total Boolean function $g$ and any constants $\epsilon, \epsilon' \in (0, 1)$. Thus, the corollary follows from Theorem 5.1 by taking $k = \lfloor 0.01n\rfloor$ and a sufficiently small constant $\epsilon > 0$. ∎

In view of the relationship between query complexity and polynomial approximation (Theorem 2.8), Corollary 5.2 gives an XOR lemma for the polynomial method in quantum query complexity:

THEOREM 5.3. *Fix (possibly partial) Boolean functions $f_1, f_2, \ldots, f_n$ on $\{-1, +1\}^m$. Then*

$$T_{\frac{1}{2}-2^{-n}}\left(\bigotimes_{i=1}^n f_i\right) \geqslant \Omega\left(\min_{|S|=\lceil 0.99n\rceil}\left\{\sum_{i\in S}\deg_{1/3}(f_i)\right\}\right).$$

This proves the XOR result in Theorem 1.5 of the Introduction.

**5.2. Direct product theorems.** We now analyze the direct product phenomenon for polynomial approximation. Analogous to the earlier development for quantum communication, we will place a lower bound on the complexity of a $(\sigma, m)$-approximant for a given set of functions.

DEFINITION 5.4. Fix finite sets $X_1, X_2, \dots, X_n \subset \mathbb{R}^m$. For a (possibly partial) Boolean function $g_i$ on $X_i$ $(i = 1, 2, \dots, n)$, we let

$$\deg(g_1, g_2, \dots, g_n, \sigma, m) = \min_{\{\phi_z\}} \max_{z \in \{-1,+1\}^n} \deg \phi_z,$$

where the minimum is over all $(\sigma, m)$-approximants $\{\phi_z\}$ of $(g_1, g_2, \dots, g_n)$.

The relevance of this definition to quantum query complexity is straightforward:

PROPOSITION 5.5. *For arbitrary* (*possibly partial*) *Boolean functions* $g_1, g_2, \dots, g_n$ *on* $\{-1, +1\}^m$,

$$T_{1-\sigma,m}(g_1, g_2, \dots, g_n) \geqslant \frac{1}{2} \deg(g_1, g_2, \dots, g_n, \sigma, m).$$

*Proof.* Given any query algorithm $A$ with cost $T$ that solves with probability $\sigma$ at least $n - m$ of the problems $g_1, g_2, \dots, g_n$, define $\phi_z(x_1, \dots, x_n) = \mathbf{P}[A(x_1, \dots, x_n) = z]$, where the probability is taken over the operation of the algorithm on a fixed input. Then $\{\phi_z\}$ is a $(\sigma, m)$-approximant for $(g_1, g_2, \dots, g_n)$. Each $\phi_z$ is the function of acceptance probabilities of a quantum query algorithm with one-bit output and cost $T$ (namely, the algorithm that accepts if and only if $A$ outputs $z$). Thus, each $\phi_z$ is a real polynomial of degree at most $2T$, by Theorem 2.8. ∎

We have:

THEOREM 5.6. *Fix a* (*possibly partial*) *Boolean function* $g_i$ *on* $X_i$ $(i = 1, 2, \dots, n)$, *for some finite sets* $X_1, X_2, \dots, X_n \subset \mathbb{R}^m$. *Then for every* $\epsilon \in (0, 1)$ *and all integers* $k, \ell, m \geqslant 0$ *with* $k + \ell \leqslant n$,

$$\deg\left(g_1, g_2, \dots, g_n, 2\binom{n}{k+1}\frac{\left(\frac{1}{2}\epsilon\right)^{k+1}}{\left(1-\frac{1}{2}\epsilon\right)^n} + 2^{-\frac{\alpha\ell^2}{n}+m+1}, m\right) \geqslant \min_{|S|=n-k-\ell}\left\{\sum_{i\in S}\deg_{1-\epsilon}(g_i)\right\},$$

*where* $\alpha > 0$ *is the absolute constant from Theorem* 3.5.

*Proof.* For each $i = 1, 2, \dots, n$, Theorem 2.4 provides $\psi_i \colon X_i \to \mathbb{R}$ that obeys (3.5), (3.6), and (5.1) for every polynomial $p$ of degree less than $\deg_{1-\epsilon}(g_i)$. Fix a $(\sigma, m)$-approximant $\{\phi_z\}$ for $(g_1, g_2, \dots, g_n)$ with $\deg \phi_z < \min_{|S|=n-k-\ell}\{\sum_{i\in S}\deg_{1-\epsilon}(g_i)\}$ for all $z$. Define functions $\Psi_k, \Phi_\ell \colon \prod X_i \to \mathbb{R}$ as in Lemma 3.7. Then $\langle \Phi_\ell, \Psi_k\rangle = 0$ by (5.1) and the assumption on the degrees of all $\phi_z$. By (3.20),

$$\sigma < 2\binom{n}{k+1}\frac{\left(\frac{1}{2}\epsilon\right)^{k+1}}{\left(1-\frac{1}{2}\epsilon\right)^n} + 2^{-\frac{\alpha\ell^2}{n}+m+1}. \qquad ∎$$

COROLLARY 5.7. *Fix a* (*possibly partial*) *Boolean function* $g_i$ *on* $X_i$ $(i = 1, 2, \dots, n)$, *for some finite sets* $X_1, X_2, \dots, X_n \subset \mathbb{R}^m$. *Let* $\beta > 0$ *be a sufficiently small absolute*

*constant. Then*

$$\deg(g_1, g_2, \ldots, g_n, \beta n, 2^{-\beta n}) \geqslant \Omega\left(\min_{|S|=\lceil 0.99n\rceil}\left\{\sum_{i\in S}\deg_{1/3}(g_i)\right\}\right).$$

*Proof.* For $n$ up to any given constant, the corollary holds trivially by choosing $\beta > 0$ suitably small and noting that a $(0.99, 0)$-approximant for $(g_1, g_2, \ldots, g_n)$ gives an approximating polynomial for each of the functions $g_1, g_2, \ldots, g_n$. For $n$ larger than a certain absolute constant, the corollary follows from Theorem 5.6 by letting $k = \ell = \lfloor 0.005n\rfloor$ and $\epsilon \in (0, 1)$ a sufficiently small constant, keeping in mind (2.9). ∎

In view of Proposition 5.5, Corollary 5.7 gives the desired direct product theorems for quantum query complexity:

THEOREM 5.8. *Fix (possibly partial) Boolean functions $f_1, f_2, \ldots, f_n$ on $\{-1, +1\}^m$. Let $\beta > 0$ be a sufficiently small absolute constant. Then*

$$T_{1-2^{-\beta n},\beta n}(f_1, f_2, \ldots, f_n) \geqslant \Omega\left(\min_{|S|=\lceil 0.99n\rceil}\left\{\sum_{i\in S}\deg_{1/3}(f_i)\right\}\right).$$

This settles the direct product result in Theorem 1.5 of the Introduction. As remarked earlier in the context of quantum communication, the constant $\beta$ can be improved by providing a sharper approximant than what is given in Theorem 3.5.

**5.3. Direct sum theorems.** We close with a direct sum property for polynomial approximation. In view of the relationship between quantum query complexity and polynomial approximation (Theorem 2.8), this gives a direct sum theorem for query complexity, incomparable with the direct product theorems derived earlier. The development here closely mirrors the setting of quantum communication and is in fact shorter and simpler.

THEOREM 5.9. *Fix a (possibly partial) Boolean function $g_i$ on $X_i$ $(i = 1, 2, \ldots, n)$, for some finite sets $X_1, X_2, \ldots, X_n \subset \mathbb{R}^m$. Then*

$$\deg_{1/3}\left(\bigotimes_{i=1}^{n} g_i\right) \geqslant \Omega\left(\sum_{i=1}^{n}\deg_{1/3}(g_i)\right).$$

*Proof.* Let $a_i = \deg_{1/3}(g_i)$ and $a = \max\{a_1, a_2, \ldots, a_n\}$. Define $S_1, S_2, S_3, \ldots$ as in Proposition 4.17. For a sufficiently small constant $\beta > 0$, we have

$$\delta = 2\prod_{i=1}^{\infty}\left(1 - \beta^{i/8}\right) - 1 > \frac{1}{3},$$

whence

$$\begin{aligned}\deg_\delta\left(\bigotimes_{i=1}^n g_i\right) &\geqslant \deg_\delta\left(\bigotimes_{i:|S_i|\geqslant i/8}\ \bigotimes_{j\in S_i} g_j\right)\\ &\geqslant \sum_{i:|S_i|\geqslant i/8} \deg_{1-\beta^{|S_i|}}\left(\bigotimes_{j\in S_i} g_j\right) &&\text{by Lemma 3.9}\\ &\geqslant \Omega\left(\sum_{i:|S_i|\geqslant i/8} |S_i|\min\{a_j : j\in S_i\}\right) &&\text{by Corollary 5.2}\\ &\geqslant \Omega\left(\sum_{i=1}^n \deg_{1/3}(g_i)\right) &&\text{by Proposition 4.17.} \qquad \square\end{aligned}$$

In view of Proposition 5.5, we infer the desired direct sum theorem for quantum query complexity, stated as Theorem 1.4 in the Introduction.

THEOREM 5.10. *Fix* (*possibly partial*) *Boolean functions* $f_1, f_2, \dots, f_n$ *on* $\{-1,+1\}^m$. *Then*

$$T_{1/3}\left(\bigotimes_{i=1}^n f_i\right) \geqslant \Omega\left(\sum_{i\in S}\deg_{1/3}(f_i)\right).$$

## 6. GENERALIZATION TO COMPOSED FUNCTIONS

In this section, we study the direct product problem in the broader context of polynomial approximation. Here, one is given Boolean functions $f_1, f_2, \dots, f_n$ and a combining function $F\colon\{-1,+1\}^n \to \{-1,+1\}$ and is asked to provide an approximating polynomial for the composition $F(f_1, f_2, \dots, f_n)$. A natural solution [14] is to compose suitable approximants for the functions in question: $\tilde F(\tilde f_1, \tilde f_2, \dots, \tilde f_n)$. The object of this section is to show that for various $F$, including random functions, this construction is optimal. The techniques of Section 3, in particular the polynomial construction of Lemma 3.1, will play an essential role in the proof.

For simplicity of exposition, we will focus here on *total* Boolean functions, although the proofs carry over readily to partial functions. The general result that will yield our sought consequences on polynomial approximation is as follows.

THEOREM 6.1. *Fix nonconstant functions* $F\colon\{-1,+1\}^n \to \{-1,+1\}$ *and* $f_i\colon X_i \to \{-1,+1\}$, $i = 1,2,\dots,n$, *for some finite sets* $X_1, X_2, \dots, X_n \subset \mathbb{R}^m$. *Then for every* $\epsilon, \delta \in (0,1)$ *and every even integer* $k \geqslant 0$,

$$\deg_{\delta-\binom{n}{k+1}\frac{2\epsilon^{k+1}}{(1-\epsilon)^n}}(F(f_1,\dots,f_n)) \geqslant \min_{|S|=\deg_\delta(F)-k}\left\{\sum_{i\in S}\deg_{1-\epsilon}(f_i)\right\}.$$

*Proof.* Put $D = \deg_\delta(F)$. Since $F, f_1, f_2, \dots, f_n$ are nonconstant, we have $D \geqslant 1$ and $\deg_{1-\epsilon}(f_i) \geqslant 1$ for all $i$. By Theorem 2.4, there exists a function $\Psi\colon\{-1,+1\}^n \to \mathbb{R}$ such

that

$$\|\Psi\|_1 = 1, \tag{6.1}$$

$$\langle \Psi, F\rangle > \delta, \tag{6.2}$$

$$\hat{\Psi}(S) = 0, \qquad |S| < D. \tag{6.3}$$

Analogously, there exist functions $\psi_i\colon X_i \to \mathbb{R}$, $i = 1, 2, \ldots, n$, such that

$$\|\psi_i\|_1 = 1,$$

$$\langle \psi_i, f_i\rangle > 1-\epsilon, \tag{6.4}$$

$$\sum_{x_i \in X_i} \psi_i(x_i)p(x_i) = 0 \tag{6.5}$$

for every polynomial $p$ of degree less than $\deg_{1-\epsilon}(f_i)$. Let $\mu$ be the product distribution on $\prod X_i$ given by $\mu(\ldots, x_i, \ldots) = \prod |\psi_i(x_i)|$. Put

$$\epsilon_{i,+1} = \mathop{\mathbf{P}}_{\mu}[f_i(x_i) \neq \operatorname{sgn}\psi_i(x_i) \mid \psi_i(x_i) > 0],$$

$$\epsilon_{i,-1} = \mathop{\mathbf{P}}_{\mu}[f_i(x_i) \neq \operatorname{sgn}\psi_i(x_i) \mid \psi_i(x_i) < 0].$$

It is clear from (6.5) that $\mathbf{P}_\mu[\psi_i(x_i) > 0] = \mathbf{P}_\mu[\psi_i(x_i) < 0] = \frac{1}{2}$ for all $i$, whence (6.4) gives $\frac{1}{2}(1-2\epsilon_{i,+1}) + \frac{1}{2}(1-2\epsilon_{i,-1}) > 1-\epsilon$ and in particular

$$\max\{\epsilon_{i,+1}, \epsilon_{i,-1}\} < \epsilon.$$

Define $\alpha_i\colon X_i \to [-1,1]$, $i = 1, 2, \ldots, n$, by

$$\alpha_i(x_i) = \begin{cases} (1-2\epsilon+\epsilon_{i,+1})/(1-\epsilon_{i,+1}) & \text{if } f_i(x_i) = \operatorname{sgn}\psi_i(x_i) = +1,\\ (1-2\epsilon+\epsilon_{i,-1})/(1-\epsilon_{i,-1}) & \text{if } f_i(x_i) = \operatorname{sgn}\psi_i(x_i) = -1,\\ -1 & \text{otherwise.}\end{cases}$$

For $z \in \{-1,+1\}^n$, let $\mu_z$ denote the probability distribution induced by $\mu$ on the set of tuples $(\ldots, x_i, \ldots)$ with $\operatorname{sgn}\psi_i(x_i) = z_i$, $i = 1, 2, \ldots, n$. The above definition of $\alpha_1, \alpha_2, \ldots, \alpha_n$ serves to ensure that $\mathbf{E}_{\mu_z}[\alpha_i(x_i)] = 1-2\epsilon$ for all $z \in \{-1,+1\}^n$ and $i = 1, 2, \ldots, n$.

Fix an even integer $k \geqslant 0$ and let $p_k\colon [-1,1]^n \to [0,\infty)$ be the degree-$k$ multilinear polynomial given by Lemma 3.1. By the above property of $\alpha_1, \alpha_2, \ldots, \alpha_n$ and multilinearity,

$$\mathop{\mathbf{E}}_{\mu_z}[p_k(\ldots, \alpha_i(x_i), \ldots)] = p_k(\ldots, 1-2\epsilon, \ldots), \qquad z \in \{-1,+1\}^n. \tag{6.6}$$

Consider the function $\zeta\colon \prod X_i \to \mathbb{R}$ given by

$$\zeta(\ldots, x_i, \ldots) = \Psi(\ldots, \operatorname{sgn}\psi_i(x_i), \ldots)\, p_k(\ldots, \alpha_i(x_i), \ldots) \prod_{i=1}^n |\psi_i(x_i)|.$$

It follows from (6.3) and (6.5) that

$$\sum_{X_1\times\cdots\times X_n} \zeta(\ldots, x_i, \ldots)\, p(\ldots, x_i, \ldots) = 0 \tag{6.7}$$

for every polynomial $p$ of degree less than $\min_{|S|=D-k}\{\sum_{i\in S} \deg_{1-\epsilon}(f_i)\}$.

CLAIM 6.2. $\|\zeta\|_1 = 2^{-n} p_k(\dots, 1-2\epsilon, \dots)$.

CLAIM 6.3.

$$\sum_{X_1\times\cdots\times X_n} \zeta(\dots, x_i, \dots) F(\dots, f_i(x_i), \dots)$$
$$> 2^{-n} p_k(\dots, 1-2\epsilon, \dots) \left\{ \delta - \frac{2\epsilon^{k+1}}{(1-\epsilon)^n} \binom{n}{k+1} \right\}.$$

In view of (6.7) and Claims 6.2 and 6.3, the proof is complete by Theorem 2.4. ∎

We now prove the required claims.

*Proof of Claim* 6.2. As mentioned earlier, $\mathbf{P}_\mu[\psi_i(x_i) > 0] = \mathbf{P}_\mu[\psi_i(x_i) < 0] = \frac{1}{2}$ for all $i$, so that the string $(\dots, \operatorname{sgn} \psi_i(x_i), \dots)$ is distributed uniformly on $\{-1,+1\}^n$ as $(\dots, x_i, \dots) \sim \mu$. Therefore,

$$\begin{aligned} \|\zeta\|_1 &= 2^{-n} \sum_{z\in\{-1,+1\}^n} |\Psi(z)| \mathop{\mathbf{E}}_{\mu_z}[p_k(\dots, \alpha_i(x_i), \dots)] && \text{by nonnegativity of } p_k \\ &= 2^{-n} p_k(\dots, 1-2\epsilon, \dots) \sum_{z\in\{-1,+1\}^n} |\Psi(z)| && \text{by (6.6)} \\ &= 2^{-n} p_k(\dots, 1-2\epsilon, \dots) && \text{by (6.1).} \end{aligned}$$ ∎

*Proof of Claim* 6.3. Let $z \in \{-1,+1\}^n$ be arbitrary. By the definition of $\mu_z$, if a string $(\dots, x_i, \dots)$ is picked according to $\mu_z$, then $f_i(x_i) = z_i$ with probability exactly $1-\epsilon_{i,z_i}$, independently for each $i$. Letting $\nu = \Pi(\dots, (\epsilon - \epsilon_{i,z_i})/(1-\epsilon_{i,z_i}), \dots)$ gives:

$$\begin{aligned} &\mathop{\mathbf{E}}_{\mu_z}[p_k(\dots, \alpha_i(x_i), \dots)\mathbf{I}[(\dots, f_i(x_i), \dots) = z]] \\ &\qquad = p_k\left(\dots, \frac{1-2\epsilon+\epsilon_{i,z_i}}{1-\epsilon_{i,z_i}}, \dots\right) \prod_{i=1}^n (1-\epsilon_{i,z_i}) \\ &\qquad = \mathop{\mathbf{E}}_{w\in\nu}[p_k(w)] \prod_{i=1}^n (1-\epsilon_{i,z_i}) && \text{by multilinearity of } p_k \\ &\qquad \geqslant \nu(1^n) p_k(1^n) \prod_{i=1}^n (1-\epsilon_{i,z_i}) && \text{by nonnegativity of } p_k \\ &\qquad = (1-\epsilon)^n p_k(1^n). \end{aligned} \tag{6.8}$$

Now,

$$
\begin{aligned}
&\left|\mathop{\mathbf{E}}_{\mu_z}[p_k(\dots,\alpha_i(x_i),\dots)F(\dots,f_i(x_i),\dots)] - p_k(\dots,1-2\epsilon,\dots)F(z)\right| \\
&\qquad = \left|\mathop{\mathbf{E}}_{\mu_z}[p_k(\dots,\alpha_i(x_i),\dots)\{F(\dots,f_i(x_i),\dots) - F(z)\}]\right| \qquad \text{by (6.6)} \\
&\qquad \leqslant 2\mathop{\mathbf{E}}_{\mu_z}[p_k(\dots,\alpha_i(x_i),\dots)(1-\mathbf{I}[(\dots,f_i(x_i),\dots)=z])] \\
&\qquad = 2\mathop{\mathbf{E}}_{\mu_z}[p_k(\dots,\alpha_i(x_i),\dots)] \\
&\qquad\qquad - 2\mathop{\mathbf{E}}_{\mu_z}[p_k(\dots,\alpha_i(x_i),\dots)\mathbf{I}[(\dots,f_i(x_i),\dots)=z]] \\
&\qquad \leqslant 2p_k(\dots,1-2\epsilon,\dots) - 2(1-\epsilon)^n p_k(1^n), \qquad (6.9)
\end{aligned}
$$

where the last step follows by (6.6) and (6.8). Since $k$ is even and $p_k$ nonnegative, we have $\mathbf{E}_{\Pi(\epsilon,\dots,\epsilon)}[|p_k(w)|] = \mathbf{E}_{\Pi(\epsilon,\dots,\epsilon)}[p_k(w)] = p_k(\dots,1-2\epsilon,\dots)$, whence Lemma 3.1 gives

$$
p_k(\dots,1-2\epsilon,\dots) \leqslant p_k(1^n)(1-\epsilon)^n\left\{1+\frac{\epsilon^{k+1}}{(1-\epsilon)^n}\binom{n}{k+1}\right\}. \qquad (6.10)
$$

We are now in a position to complete the proof of the claim. As mentioned before, the string $(\dots,\operatorname{sgn}\psi_i(x_i),\dots)$ is distributed uniformly on $\{-1,+1\}^n$ as $(\dots,x_i,\dots)\sim\mu$. Thus,

$$
\begin{aligned}
&\sum_{X_1\times\cdots\times X_n}\zeta(\dots,x_i,\dots)F(\dots,f_i(x_i),\dots) \\
&\qquad = 2^{-n}\sum_{z\in\{-1,+1\}^n}\Psi(z)\mathop{\mathbf{E}}_{\mu_z}[p_k(\dots,\alpha_i(x_i),\dots)F(\dots,f_i(x_i),\dots)] \\
&\qquad \geqslant 2^{-n}\sum_{z\in\{-1,+1\}^n}\Psi(z)F(z)p_k(\dots,1-2\epsilon,\dots) \\
&\qquad\qquad - 2^{-n}\sum_{z\in\{-1,+1\}^n}|\Psi(z)|\{2p_k(\dots,1-2\epsilon,\dots)-2(1-\epsilon)^n p_k(1^n)\} \\
&\qquad > 2^{-n}\delta p_k(\dots,1-2\epsilon,\dots) - 2^{-n}\{2p_k(\dots,1-2\epsilon,\dots)-2(1-\epsilon)^n p_k(1^n)\},
\end{aligned}
$$

where the next-to-last step uses (6.9) and the last step uses (6.1) and (6.2). In view of (6.10), the proof is complete. ∎

This completes the proof of Theorem 6.1. We will now derive several results on polynomial approximation by setting the parameters in Theorem 6.1 in various ways.

THEOREM 6.4. *Fix nonconstant functions* $F\colon\{-1,+1\}^n\to\{-1,+1\}$ *and* $f_i\colon X_i\to\{-1,+1\}$, $i=1,2,\dots,n$, *for some finite sets* $X_1,X_2,\dots,X_n\subset\mathbb{R}^m$. *Fix* $\epsilon,\delta\in(0,1)$ *with* $\deg_\delta(F)\geqslant 30\epsilon n$. *Then*

$$
\deg_{\delta-2^{-\epsilon n}}(F(f_1,\dots,f_n)) \geqslant \min_{|S|=\lceil\deg_\delta(F)/2\rceil}\left\{\sum_{i\in S}\deg_{1-\epsilon}(f_i)\right\}.
$$

*Proof.* Apply Theorem 6.1 with $k=2\lfloor 7.5\epsilon n\rfloor$. ∎

Anthony [6] and Saks [45] show that almost every function $F\colon \{-1,+1\}^n \to \{-1,+1\}$ obeys $\deg_\pm(F) \geqslant n/2$. Therefore, recalling (2.9) and (2.11) gives the following corollary to Theorem 6.4:

COROLLARY 6.5. *Fix nonconstant functions $f_i\colon X_i \to \{-1,+1\}$, $i = 1,2,\ldots,n$, for some finite sets $X_1, X_2, \ldots, X_n \subset \mathbb{R}^m$. Then*

$$\deg_{1-2^{-n/120}}(F(f_1,\ldots,f_n)) \geqslant \Omega\left(\min_{|S|=\lceil n/4\rceil}\left\{\sum_{i\in S}\deg_{1/3}(f_i)\right\}\right)$$

*for almost every function $F\colon \{-1,+1\}^n \to \{-1,+1\}$.*

In particular, the corollary shows that $\deg_{1-2^{-\Omega(n)}}(F(f,f,\ldots,f)) = \Omega(n\deg_{1/3}(f))$ for almost all functions $F\colon \{-1,+1\}^n \to \{-1,+1\}$. This bound is tight in a strong sense: Buhrman, Newman, Röhrig, and de Wolf [14] show that $\deg_{1/3}(F(f,f,\ldots,f)) = O(n\deg_{1/3}(f))$ for every $F\colon \{-1,+1\}^n \to \{-1,+1\}$. We derive one additional result, valid for all functions $F$.

THEOREM 6.6. *Fix nonconstant functions $F\colon \{-1,+1\}^n \to \{-1,+1\}$ and $f_i\colon X_i \to \{-1,+1\}$, $i = 1,2,\ldots,n$, for some finite sets $X_1, X_2, \ldots, X_n \subset \mathbb{R}^m$. Then*

$$\deg_{1/3}(F(f_1,\ldots,f_n)) \geqslant \Omega\left(\deg_{1/3}(F)\min_{i=1,\ldots,n}\left\{\deg_{1-\frac{1}{30n}\deg_{0.99}(F)}(f_i)\right\}\right) \qquad (6.11)$$

$$\geqslant \Omega\left(\frac{\deg_{1/3}(F)^2}{n}\cdot\min_{i=1,\ldots,n}\{\deg_{1/3}(f_i)\}\right). \qquad (6.12)$$

*In particular, taking $f_1 = f_2 = \cdots = f_n = f$,*

$$\deg_{1/3}(F(f,f,\ldots,f)) \geqslant \Omega\left(\frac{\deg_{1/3}(F)^2}{n}\cdot\deg_{1/3}(f)\right). \qquad (6.13)$$

*Proof.* To obtain (6.11), invoke Theorem 6.4 with $\delta = 0.99$ and $\epsilon = \deg_\delta(F)/(30n)$, and use (2.9). To obtain (6.12), apply (2.9) and (2.10). ∎

The quoted result by Buhrman et al. [14] shows that (6.13) is tight for any function $F$ with $\deg_{1/3}(F) = \Theta(n)$, including familiar functions such as majority, parity, and the random functions.

In view of the relationship between polynomial approximation and query complexity (Theorem 2.8), the results of this section immediately translate into lower bounds for quantum query complexity. More explicitly, Theorems 6.1, 6.4, 6.6, and Corollary 6.5 give lower bounds on the quantum query complexity of composed functions $F(f_1, f_2, \ldots, f_n)$ in terms of the approximate degrees of the constituent functions $F, f_1, f_2, \ldots, f_n$.

*Further extensions.* As remarked earlier, the proof of Theorem 6.1 carries over to the setting of partial Boolean functions. After minor changes, it also gives lower bounds on the communication complexity of composed problems of the form $F(F_1, F_2, \ldots, F_n)$, where $F$ is a (possibly partial) Boolean function on $\{-1,+1\}^n$ and $F_1, F_2, \ldots, F_n$ are (possibly partial) sign matrices. We defer these refinements to the final version of the paper.